\documentclass{llncs}

\usepackage{mystyle}

\begin{document}

\title{Approximating Local Homology from Samples}

\author{Primoz Skraba\inst{1} \and 
Bei Wang\inst{2}  
}

\institute{Jo\v zef Stefan Institute, Slovenia\and
Scientific Computing and Imaging Institute, University of Utah, USA
}

\maketitle


\begin{abstract}

\vspace{-2mm}
Recently, multi-scale notions of local homology (a variant of
persistent homology) have been used to study the local structure
of spaces around a given point from a point cloud sample. Current
reconstruction guarantees rely on constructing embedded complexes
which become difficult in high dimensions. We show
that the persistence diagrams used for estimating local homology,
can be approximated using families of Vietoris-Rips complexes,
whose simple constructions are robust in any dimension.  To the best of
our knowledge, our results, for the first time, make applications based on local homology, such as stratification learning, feasible in
high dimensions. 

\end{abstract}



\section{Introduction}
\vspace{-2mm}

Advances in scientific and computational experiments have
improved our ability to gather large collections of data points
in high-dimensional spaces. One aspect in topological data
analysis is  to infer the topological structure of a space 
given a point cloud sample.  
We often assume the space has manifold structure, however, 
more interesting cases arise when we relax our assumptions to include 
spaces that contains singularities and mixed dimensionality, for example, 
stratified spaces.

Stratified spaces can be decomposed into
manifold pieces that are glued together in some uniform way. 
An important tool in studying these spaces is the
study of the neighborhoods surrounding singularities, where
manifolds of different dimensionality intersect. 
 We focus on sampling conditions for such neighborhoods, which
allow us to begin examining how difficult certain reconstruction
techniques are with respect to the geometric properties of the
underlying shape.  
Our main task is to infer sampling conditions suitable
for recovering local structures of stratified spaces, in
particular, the \emph{local homology groups}, from a possibly
noisy sampled point set.

\noindent\textbf{Stratification learning.} 
In stratification learning (or mixed manifold learning), a
point cloud is assumed to be sampled from a mixture of (possibly
intersecting) manifolds. The objective is to recover the
different pieces, often treated as clusters, of the data
associated with different manifolds of varying dimensions.
Stratified spaces has been studied extensively in mathematics,
see seminal work in~\cite{GorMac1988,Wei1994}.  Recently,
topological data analysis, relying heavily on ingredients from
computational topology
\cite{EdeCohZom2002,ZomorodianCarlsson2005} and intersection
homology \cite{GoreskyMacPherson1982,Ben2008,BendichHarer2011}
has gained momentum in stratification learning.  In particular,
the work in \cite{BenCohEde2007} focuses on studying the local
structure of a sampled stratified spaces based on a multi-scale
notion of local homology (see Section \ref{sec:background}).
More recent work in \cite{BendichWangMukherjee2012} studies how
point cloud data could be clustered by strata based on how the
local homology of nearby sampled points map into one another.

\noindent\textbf{Reconstruction and sampling.}  Reconstructing
shapes from potential noisy point cloud samples has been studied
in many fields.  Most often the work is heavily tied to a
reconstruction criteria (e.g. homotopic, homeomorphic, etc.) and
the assumptions on the underlying space (e.g. manifold).
Combinatorial algorithms in geometry are generally derived from
Delaunay triangulations \cite{CazalsGiesen2006} and alpha shapes
\cite{EdelsbrunnerMucke1992}, and provide correctness proofs
associated with such reconstructions \cite{Dey2007}.  As the
dimension increases, reconstruction efforts have been redirected
towards alternative combinatorial structures such as tangential
Delaunay complexes \cite{BoissonnatGhosh2010}, witness complexes
\cite{SilCar2004}, \v{C}ech complexes and the closely related
Vietoris-Rips complexes
\cite{ChazalOudot2008,AttLieSal2011b,AttLieSal2011}.

However, these existing techniques are primarily concerned with global
reconstruction. Providing reconstruction guarantees for local
structures is more challenging.  To guarantee theoretical
correctness in computing persistence local homology, both
\cite{BenCohEde2007} and \cite{BendichWangMukherjee2012} use
Delaunay complexes and their variants. However constructing Delaunay
complexes in high dimensions is known to be difficult due to
scaling and numerical issues with predicates.  
On the other hand, methods for
fast \cite{Zom2010} and efficient
\cite{AttLieSal2011b,Sheehy2012} constructions of Vietoris-Rips
complexes are available, and there have been theoretical advances
on their topology-preserving qualities, 
making it appealing for computations in high dimensions. The goal of
this paper is to make persistent local homology computation more
practical through approximations.

\noindent\textbf{Contributions.}  Our contributions focus on providing 
sampling conditions to recover the local structure of a
space from a point cloud sample, based on previously
introduced \cite{BenCohEde2007} multi-scale notions of local
homology.  Our main results are:
\vspace{-5mm}
\begin{itemize} \denselist
\item We extend previously introduced algebraic
  constructions in the analysis of scalar fields over point cloud data \cite{ChazalGuibasOudot2009} to two multi-scale notions of local homology. 
\item For both multi-scale notions of local homology, we approximate 
their persistence diagrams by constructing families of Vietoris-Rips complexes 
based on a set of sample points, formalized within Theorem \ref{theorem:alpha-sampling} and \ref{theorem:r}.     
  The simplicity and efficiency of building the
  these complexes in any dimension makes, 
  for the first time,
 applications based on local homology such as  stratification learning feasible in
  high dimensions.
\item We show that relative persistent modules are interleaved if
  the respective absolute persistent modules are
  interleaved. We consider such a technical result (Theorem \ref{theorem:quotient_interleave}) of
  independent interest.
\item Our results imply algorithms for computing the local
  homology either by a reduction to standard persistence or a
  known variant.
\end{itemize}
\vspace{-5mm}


\section{Background}
\label{sec:background}

The background material focuses on the introduction of
persistence modules \cite{ChaCohGli2009}, local homology and its
multi-scale notions \cite{BenCohEde2007}. 
We assume a basic knowledge of homology and
persistent homology, see \cite{Mun1984,Hat2002} for a readable
background of the former, and \cite{EdeHar2010} for a
computational treatment of the latter.

\noindent\textbf{Persistence Modules.}  We use the definition of
persistence modules adapted from \cite{ChaCohGli2009}.  A
\emph{persistence module} $\Fcal = \{F_{\alpha}\}_{\alpha
  \in \Rspace}$ is a collection of vector spaces $F_{\alpha}$
(over any fields) together with a family $\{f_{\alpha}^{\beta}:
F_{\alpha} \to F_{\beta}\}_{\alpha \leq \beta}$ of linear maps
such that $\alpha \leq \beta \leq \gamma$ implies
$f_{\alpha}^{\gamma} = f_{\beta}^{\gamma} \circ
f_{\alpha}^{\beta}$, and $f_{\alpha}^{\alpha} = id_{F_{\alpha}}$.
A persistence module is \emph{tame} if it has finite number of
critical values and all $F_a$ are of finite rank.  Unless
otherwise specified, we suppose all persistence modules we
encounter in this paper are tame. 

\begin{figure}[!t]
\begin{center}
  \includegraphics[width=1.0\textwidth]{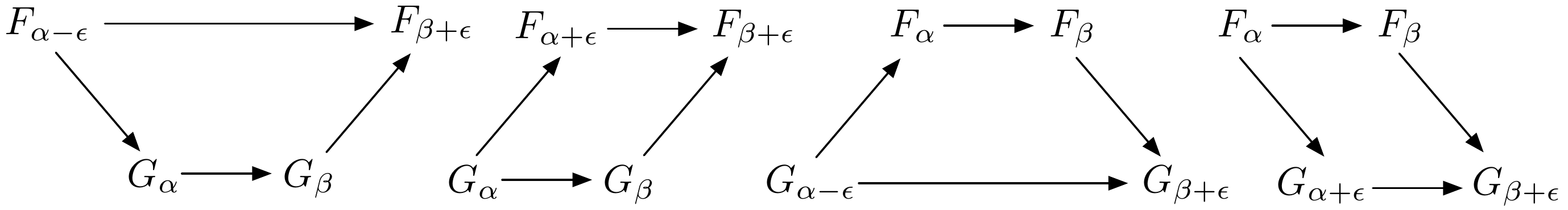}
\end{center}
\vspace{-5mm}
  \caption{Strongly $\ep$-interleaved persistence modules.}
  \label{fig:epinterleave}
\end{figure}

Two persistence modules $\{F\}_{\alpha}$ and $\{G\}_{\alpha}$ are
\emph{(strongly) $\ep$-interleaved} if there exists two families
of homomorphisms, $\mu_{\alpha}: F_{\alpha} \to G_{\alpha + \ep}$
and $\nu_{\alpha}: G_{\alpha} \to F_{\alpha + \ep}$, that make
the following diagrams (Fig. \ref{fig:epinterleave}) commute
for all $\alpha \leq \beta \in \Rspace$ \cite{ChaCohGli2009}.
The information contained in a persistence module can be encoded
by a multi-set of points in the extended plane $\bar{\Rspace}^2$ 
(where $\bar{\Rspace} = \Rspace \cup \{-\infty, \infty\}$),
called a \emph{persistence diagram}  \cite{CohEdeHar2007}.  If two
tame persistence modules are $\ep$-interleaved, the bottleneck
distance between their persistence diagrams are upper bounded by
$\ep$ (\cite{CohEdeHar2007},Theorem 4.4).
In this paper, we consider persistence modules of homology groups and relative homology groups over a field. 
Given a family of topological spaces $\{\Xspace_{\alpha}\}_{\alpha}$ connected by inclusions 
$\Xspace_{\alpha} \hookrightarrow \Xspace_\beta,$
the inclusions induce a sequence of homology groups connected by homomorphisms, $\Hgroup_k(\Xspace_{\alpha}) \to \Hgroup_k({\Xspace_\beta})$, where $k$ is the homological dimension. 
We therefore obtain persistence modules of the form 
$\{\Hgroup_k(\Xspace_{\alpha})\}_{\alpha}$. 
Specifically, when the linear maps associated with two persistence modules $\{\Hgroup_k(\Xspace_{\alpha})\}_{\alpha}$ and $\{\Hgroup_k(\Yspace_{\alpha})\}_{\alpha}$ are induced by inclusions at the space level $\Xspace_{\alpha} \hookrightarrow \Yspace_{\alpha+\ep}$ 
and $\Yspace_{\alpha} \hookrightarrow \Xspace_{\alpha+\ep}$, 
their $k$-th persistence modules are $\ep$-interleaved \cite{ChaCohGli2009}. 
For the rest of the paper, we sometimes abuse this notation by omitting the $k$-th homology functor unless necessary.
We work with singular homology here but our results are applicable in the simplicial setting as well. 

\begin{figure}[!t]
\begin{center}
  \def\svgwidth{0.4\columnwidth} 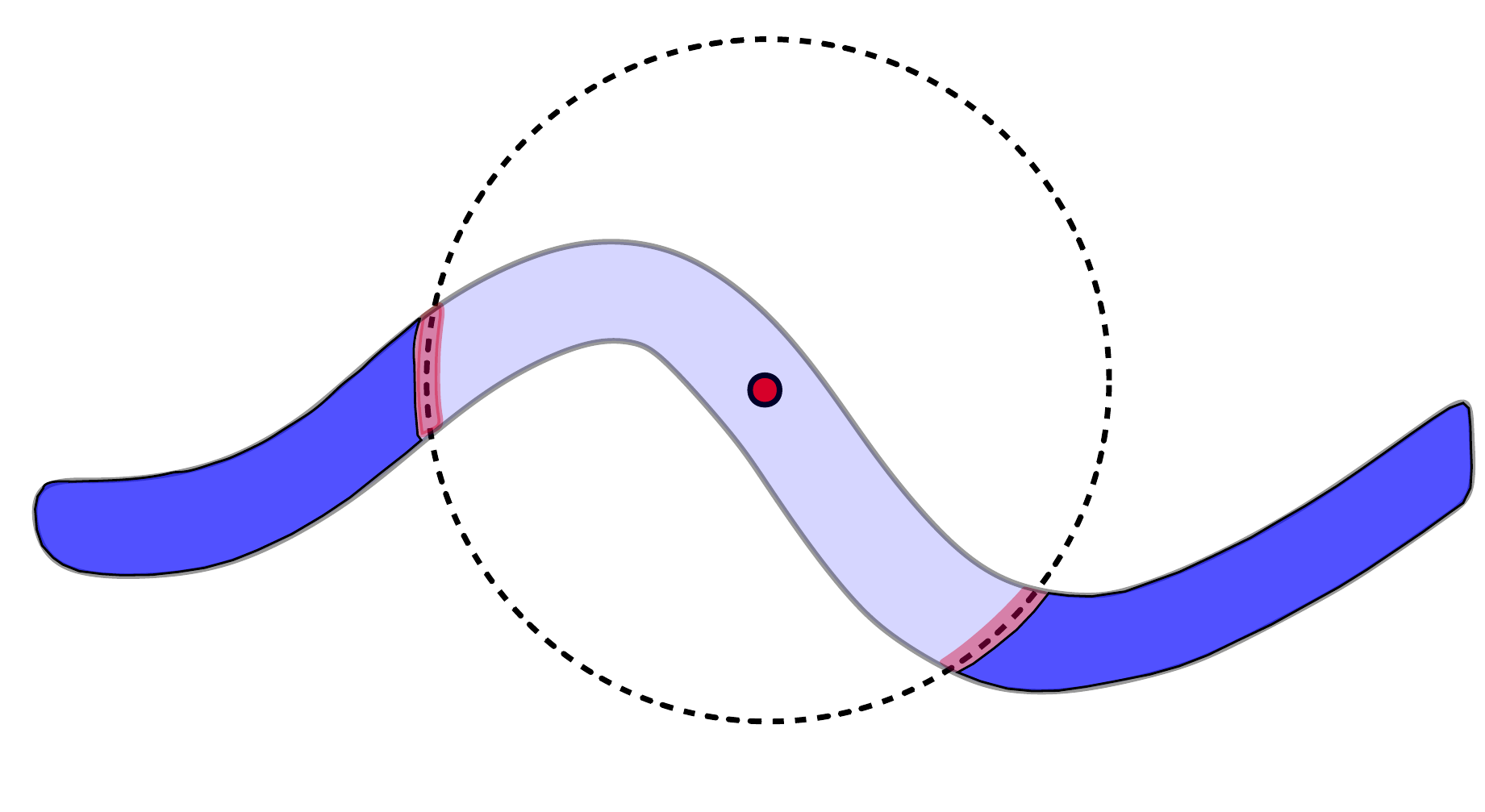
  \def\svgwidth{0.4\columnwidth} 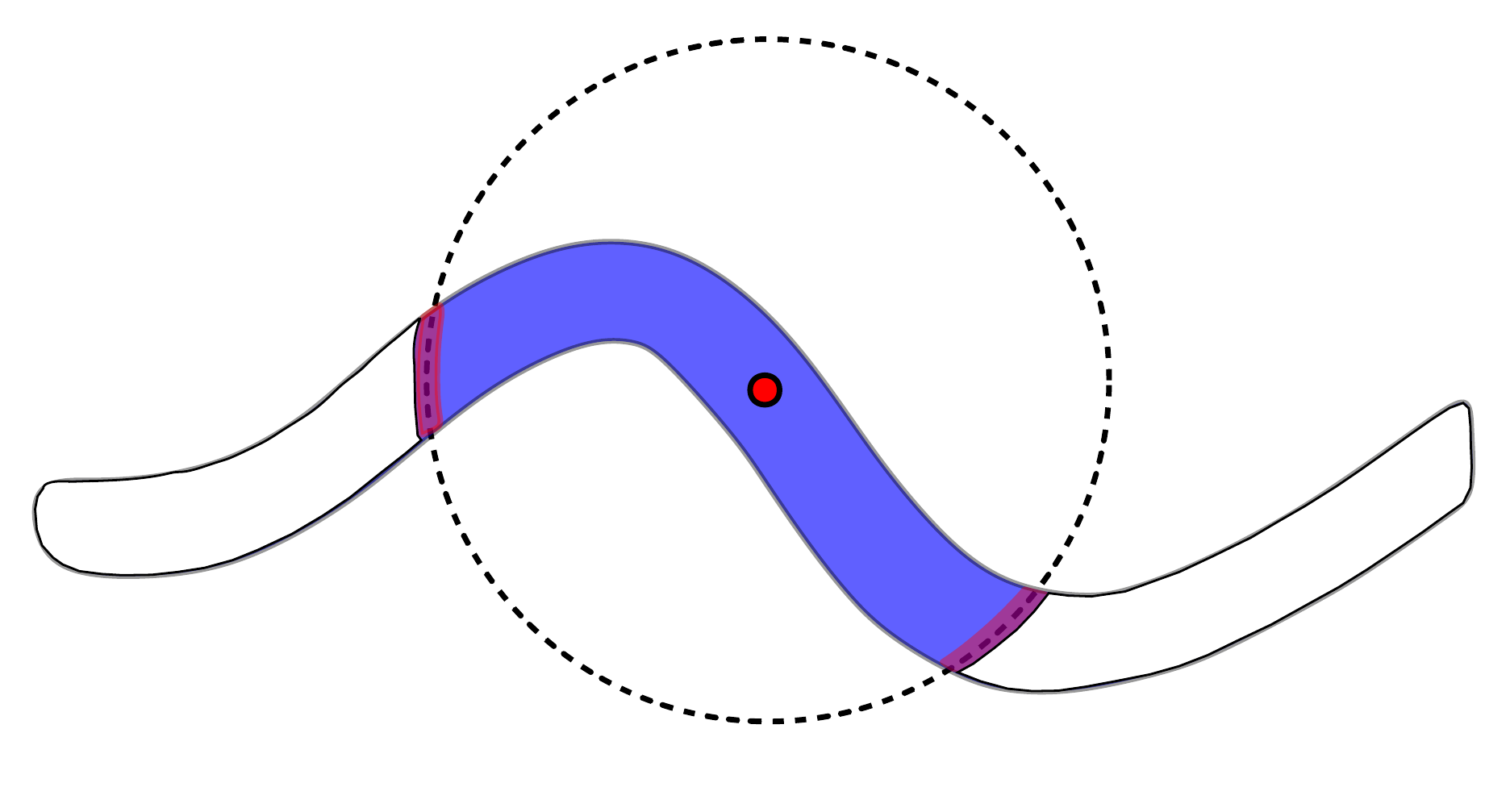
\end{center}
\vspace{-7mm}
  \caption{Local homology as the direct limit, $\lim_{r \to 0}\Hgroup(\Xspace, \Xspace \cap B^r)$ (left) or $\lim_{r \to 0}\Hgroup(\Xspace \cap B_r, \Xspace \cap \bdr B_r)$ (right).}
  \label{fig:localhomology}
\vspace{-5mm}
\end{figure}

\noindent\textbf{Local Homology.}
The \emph{local homology groups} at a point $x \in \Xspace$ is
defined as the relative homology groups $\Hgroup(\Xspace, \Xspace
- x)$ (\cite{Mun1984}, page 126). In this paper, we assume that the topological space $\Xspace$ is
embedded in some Euclidean space $\Rspace^d$ \footnote{This assumption can be
  relaxed in several ways but this setting is most common in our
  applications.}.
Let $d_x: \Rspace^d \to \Rspace$ be the Euclidean distance
function from a fixed $x \in \Xspace$, $d_x(y) := d(x,y) = ||y - x||$.  Let $B_r
= B_r(x) = d_x^{-1}[0, r]$ and $B^r = B^r(x) = d_x^{-1}[r,
  \infty)$ be the sublevel sets and superlevel sets of $d_x$.
  Taking a small enough $r$, the local homology groups in
  questions are in fact the \emph{direct limit} of relative
  homology groups, $\lim_{r \to 0}\Hgroup(\Xspace, \Xspace \cap
  B^r)$, or alternatively $\lim_{r \to 0}\Hgroup(\Xspace \cap
  B_r, \Xspace \cap \bdr B_r)$\cite{Ben2008}, see Fig. 
  \ref{fig:localhomology}.  We then adapt two multi-scale
  notions of this concept based on persistence (which are first introduced in \cite{BenCohEde2007}),  
  referred to as the \emph{$r$-filtration} and the \emph{$\alpha$-filtration}.
  The goal of this paper is to derive sampling conditions that are
appropriate to compute the persistence diagrams with respect to
these filtrations, therefore approximating the local homology at $x
\in \Xspace$.

For a fixed $\alpha \geq 0$, let $\Xspace_{\alpha}$ be the ``thickened'' or ``offset" version of $\Xspace$, that is, the space of points in $\Rspace^d$ at Euclidean distance at most $\alpha$ from $\Xspace$.
Suppose $L$ is a finite set of points sampled from $\Xspace$ \footnote{Our results would hold with minor modifications in the setting of sampling with noise, where elements of $L$ lie on or near $\Xspace$.}, where $L \subset \Xspace$
 and $L_{\alpha} = \cup_{x \in L} B_{\alpha}(x)$. 
In subsequent sections, we put further restrictions on $L$ where we suppose $L$ is an 
\emph{$\ep$-sample} of $\Xspace$, that is, $\forall x \in \Xspace$, $d(x, L) := \inf_{y \in L} d(x,y) \leq \ep$. 

\begin{figure*}[!ht]
\vspace{-3mm}
\begin{center}
  \def\svgwidth{0.4\columnwidth} 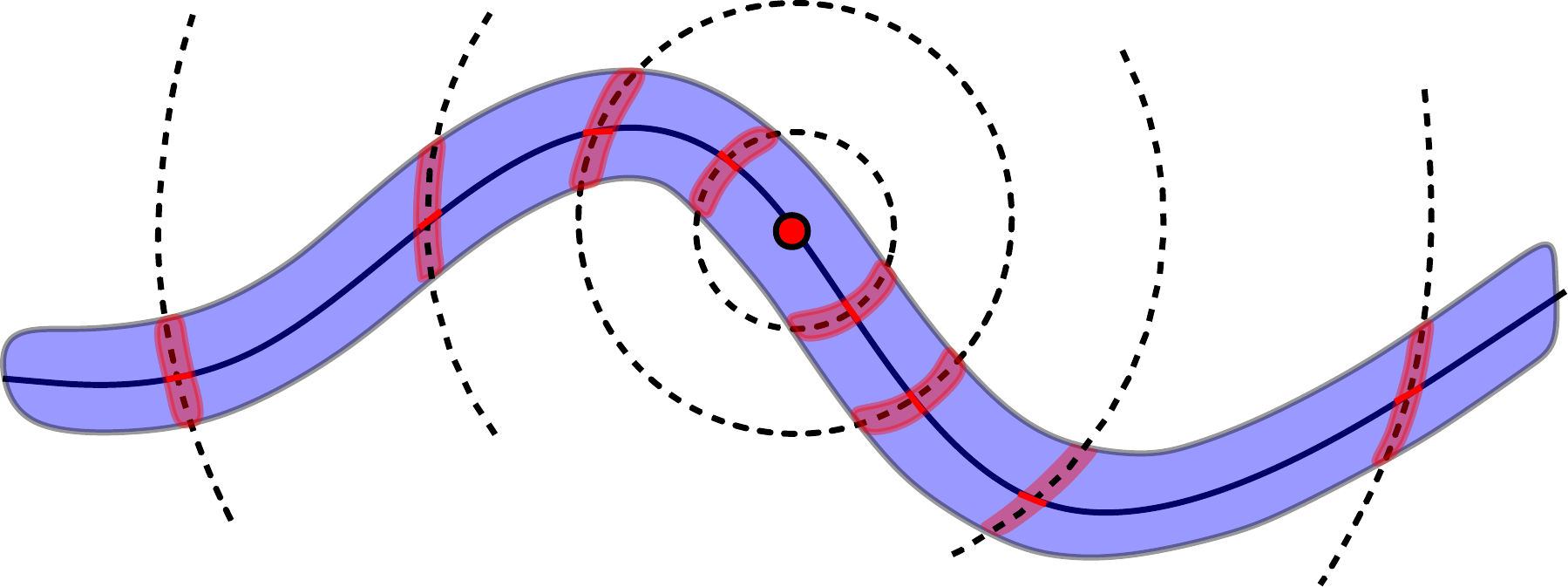
  \def\svgwidth{0.4\columnwidth} 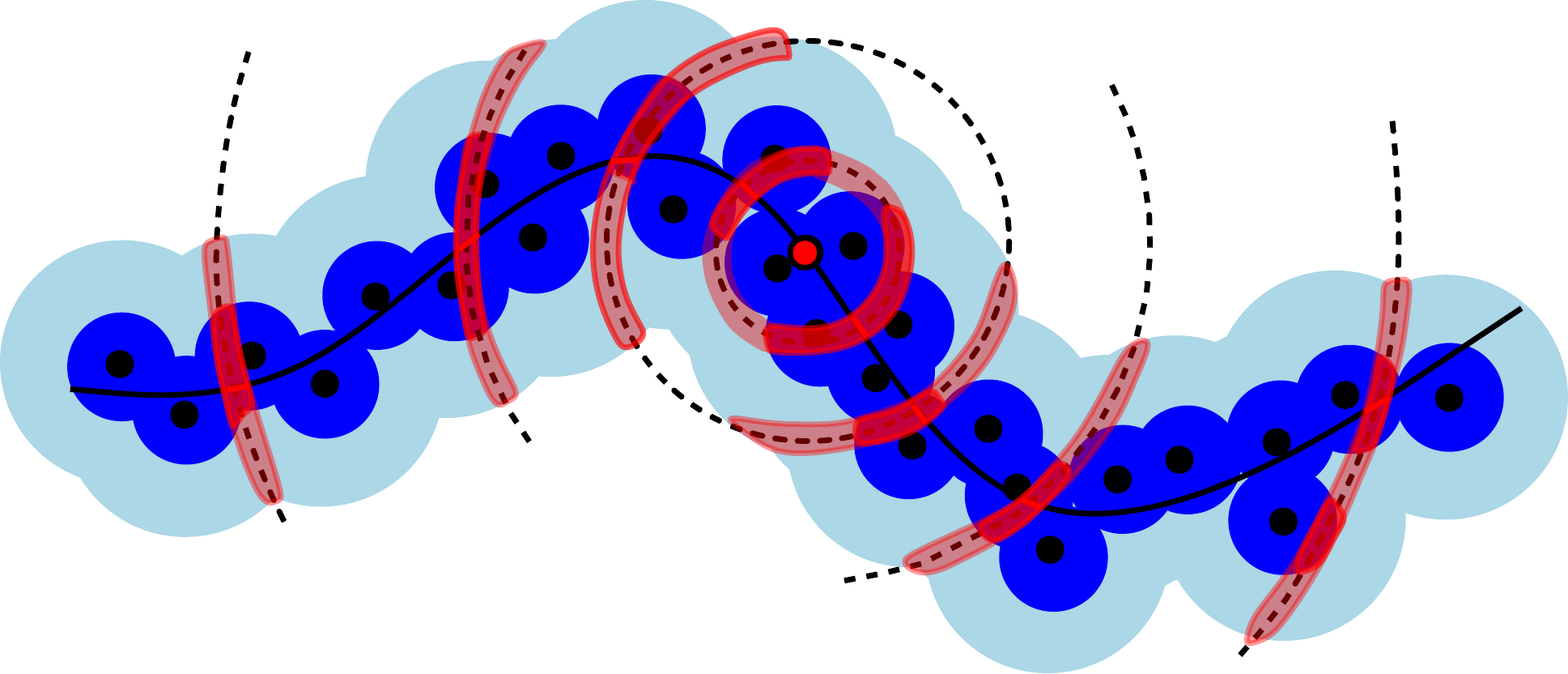
\end{center}
\vspace{-5mm}
  \caption{The $r$-filtration for space $\Xspace$ and its offsets (left), and the same
  filtration built on a set of points $L$, sampled from $\Xspace$.}
  \label{fig:r-filtration}
\end{figure*}

The \emph{$r$-filtration} (Fig. \ref{fig:r-filtration}) is a sequence
of relative homology groups connected by linear maps induced by
inclusion and excision, constructed by fixing a thickening
parameter $\alpha$ and varying parameter $r$, for $r'> r$,
$$\cdots \to \Hgroup(\Xspace_{\alpha}, \Xspace_{\alpha} \cap B^{r'}) 
\to \Hgroup(\Xspace_{\alpha}, \Xspace_{\alpha} \cap B^{r}) \to \cdots.$$
The same filtration could be built on a set of points $L$ sampled from $\Xspace$, that is, 
$$\cdots \to \Hgroup(\Lspace_{\alpha}, \Lspace_{\alpha} \cap B^r) 
\to \Hgroup(\Lspace_{\alpha}, \Lspace_{\alpha} \cap B^{r'}) \to \cdots.$$
Here, we fix the space at resolution $\alpha$, and vary the scale $r$ at which we analyze the local neighborhood, analog to changing the lens from the front of the camera. 

\begin{figure*}[!ht]
\vspace{-5mm}
\begin{center}
   \def\svgwidth{0.4\columnwidth} 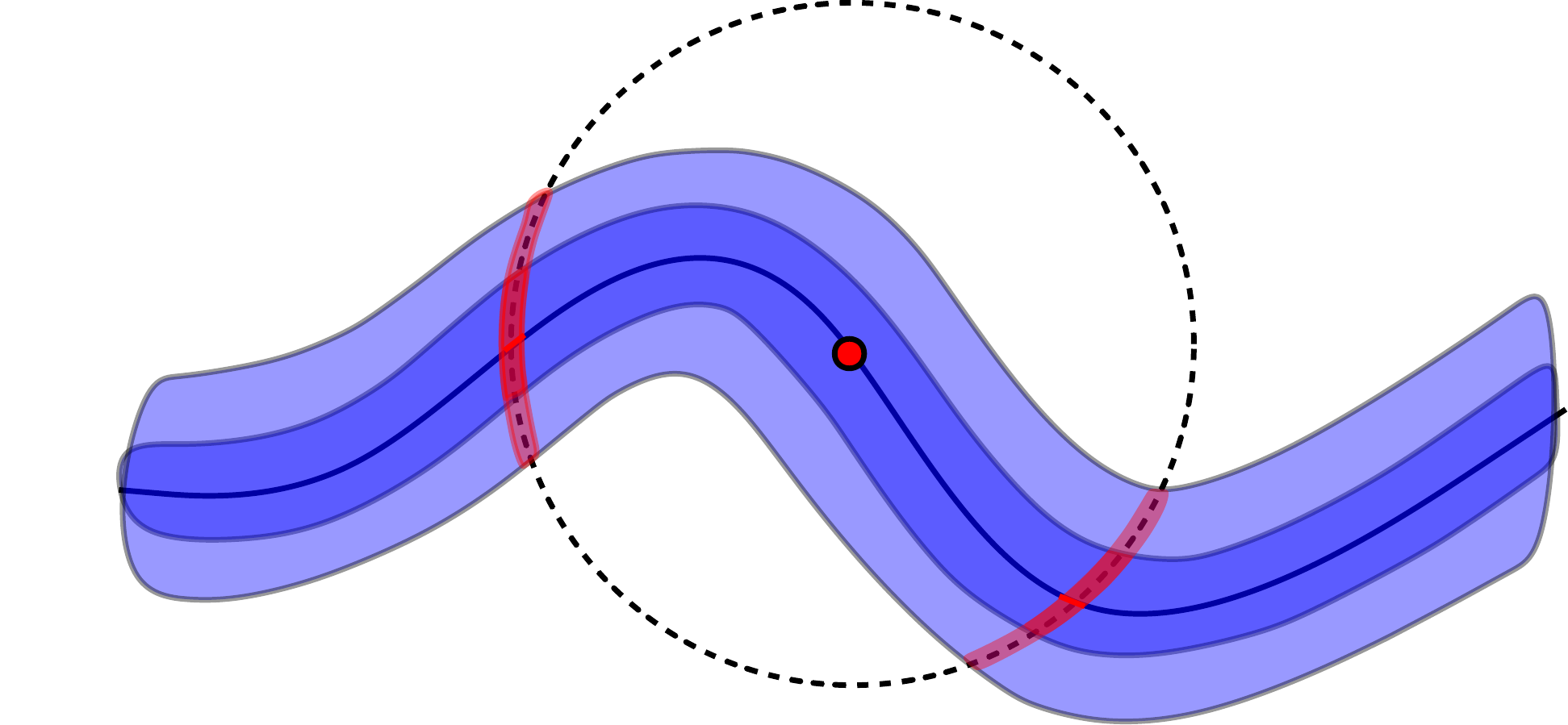
  \def\svgwidth{0.4\columnwidth} 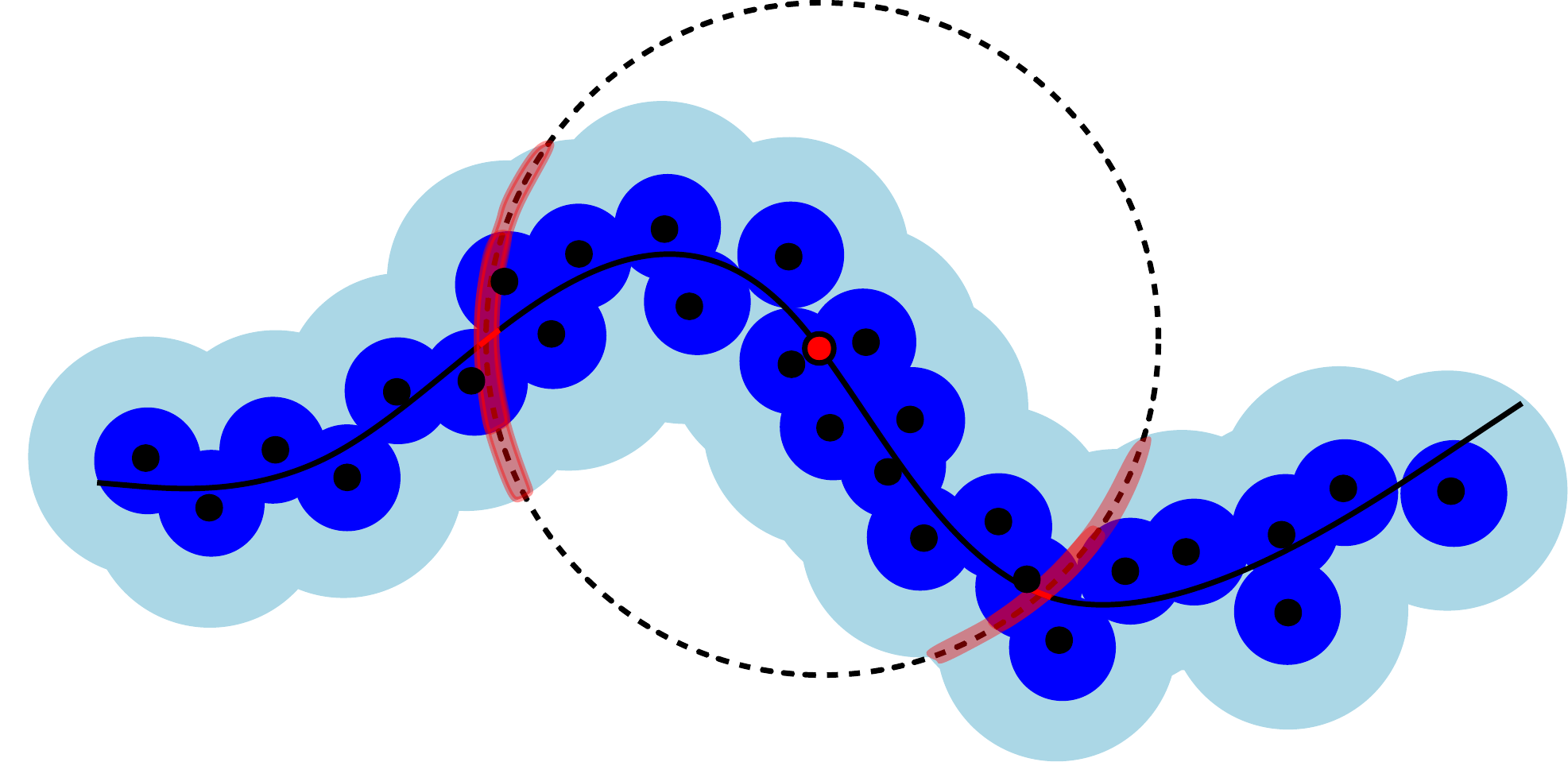
\end{center}
\vspace{-5mm}
  \caption{The $\alpha$-filtration for space $\Xspace$ and its offset (left), and on the right, the same
  filtration built on a set of points $L$, sampled from $\Xspace$.}
  \label{fig:alpha-filtration}
\end{figure*}

The \emph{$\alpha$-filtration}  (Fig. \ref{fig:alpha-filtration}) is a sequence of relative homology groups connected by inclusion,  constructed by fixing $r$ and varying $\alpha$, for $\alpha < \alpha'$,
$$\cdots \to \Hgroup(\Xspace_{\alpha} \cap B_r, \Xspace_{\alpha} \cap \bdr B_r) 
\to \Hgroup(\Xspace_{\alpha'} \cap B_r, \Xspace_{\alpha'} \cap \bdr B_{r}) \to \cdots.$$
Its discrete counterpart built on a set of points $L$ sampled from $\Xspace$ is, 
$$\cdots \to \Hgroup(\Lspace_{\alpha} \cap B_r, \Lspace_{\alpha} \cap \bdr B_r) 
\to \Hgroup(\Lspace_{\alpha'} \cap B_r, \Lspace_{\alpha'} \cap \bdr B_{r}) \to \cdots.$$
Here, we fix the size of the ball which defines the locality, 
i.e. the size $r$ of the local neighborhood, and we vary the scale $\alpha$ at which we analyze the
space.   

\noindent\textbf{\v Cech and Vietoris-Rips Complexes.}
Suppose $L$ is a finite point set in $\Rspace^d$ and 
$L_{\alpha} = \cup_{x \in L} B_{\alpha}(x)$.
The nerve of $L_{\alpha}$ is the \emph{\v Cech complex} of $L$, denoted as $\Ccal_\alpha = \Ccal_{\alpha}(L)$ (omitting $L$ from the notation unless necessary). 
The \emph{Vietoris-Rips complex} of $L$ is denoted as $\Rcal_{\alpha}$, 
whose simplicies correspond to non-empty subsets of $L$ of diameter less than $\alpha$. 
For Euclidean metric space, we have, $\forall \alpha > 0$,
$\Ccal_{\alpha/2} \subseteq \Rcal_{\alpha} \subseteq \Ccal_{\alpha} \subseteq \Rcal_{2\alpha}$.
This implies that the persistence modules $\{\Hgroup(\Ccal_{\alpha})\}_{\alpha}$ and $\{\Hgroup(\Rcal_{\alpha})\}_{\alpha}$ are $\alpha$-interleaved.


\section{Approximating Local Homology: $\alpha$-Filtration}
\label{sec:alpha}

In the $\alpha$-filtration, since we will be computing relative persistent homology, there
are certain requirements on the pairs, such that the maps of the
relative filtration are well-defined. 
Two filtrations, $\Acal = \{A_{\alpha}\}_{\alpha \in \Rspace}$ and 
$\Fcal = \{F_{\alpha}\}_{\alpha \in \Rspace}$ are called \emph{compatible}
if for all $\alpha \leq \beta$, the following diagram commutes:
\vspace{-0.4cm}
\begin{equation*}
\begin{tikzcd}
A_\alpha \arrow{r}{}\arrow{d}{} & F_\alpha \arrow{d}{}\\
A_\beta \arrow{r}{} & F_\beta. 
\end{tikzcd}
\end{equation*}
\vspace{-0.4cm}

This ensures that the relative persistence module is
well-defined \footnote{Note that this is equivalent to the condition given
on pairs of filtrations under the two function setting \cite{CohEdeHar2009b}.}. 
In our context, all the maps
are induced by inclusions hence the above diagram commutes.
We highlight steps involved to obtain our approximation results:
\begin{itemize}\denselist
\item First, we show that under certain conditions, the relative homology of a ball modulo its boundary is isomorphic to that of the entire space modulo the subspace outside the ball. 
\item Second, we prove that if we have two compatible filtrations
  $\Fcal$ and $\Acal$ which are respectively interleaved with
  $\Gcal$ and $\Bcal$, the relative persistent homology
  $\Hgroup(\Fcal , \Acal)$ is approximated by $\Hgroup(\Gcal ,
  \Bcal)$. This result may be of independent interest.
\item Last, we prove a series of inter-leavings to show that both
  filtrations in our case can be interleaved with a Vietoris-Rips 
  construction on the samples.
\end{itemize}

\begin{figure}[!t]
\begin{center}
\includegraphics[width=0.4\linewidth]{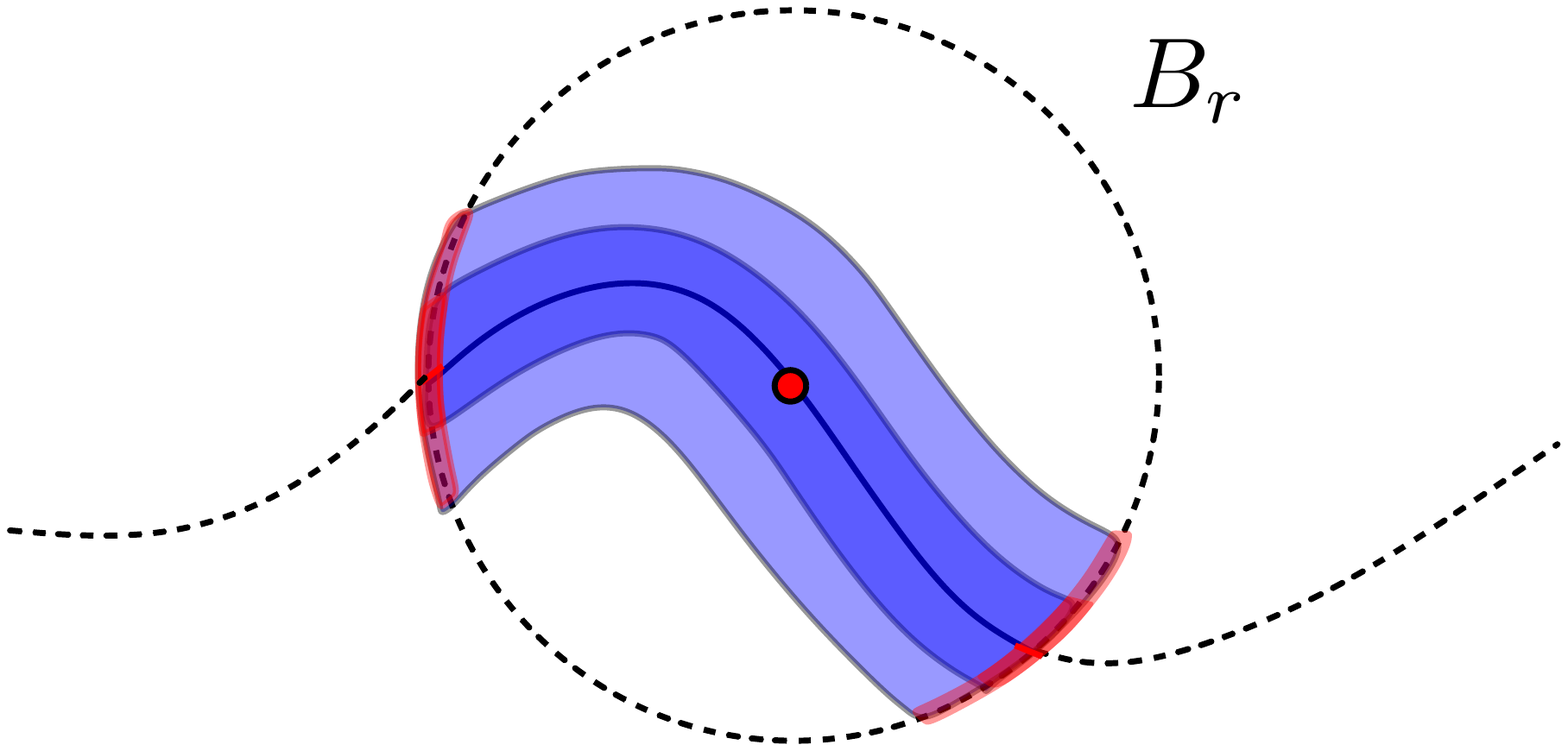} 
\includegraphics[width=0.4\linewidth]{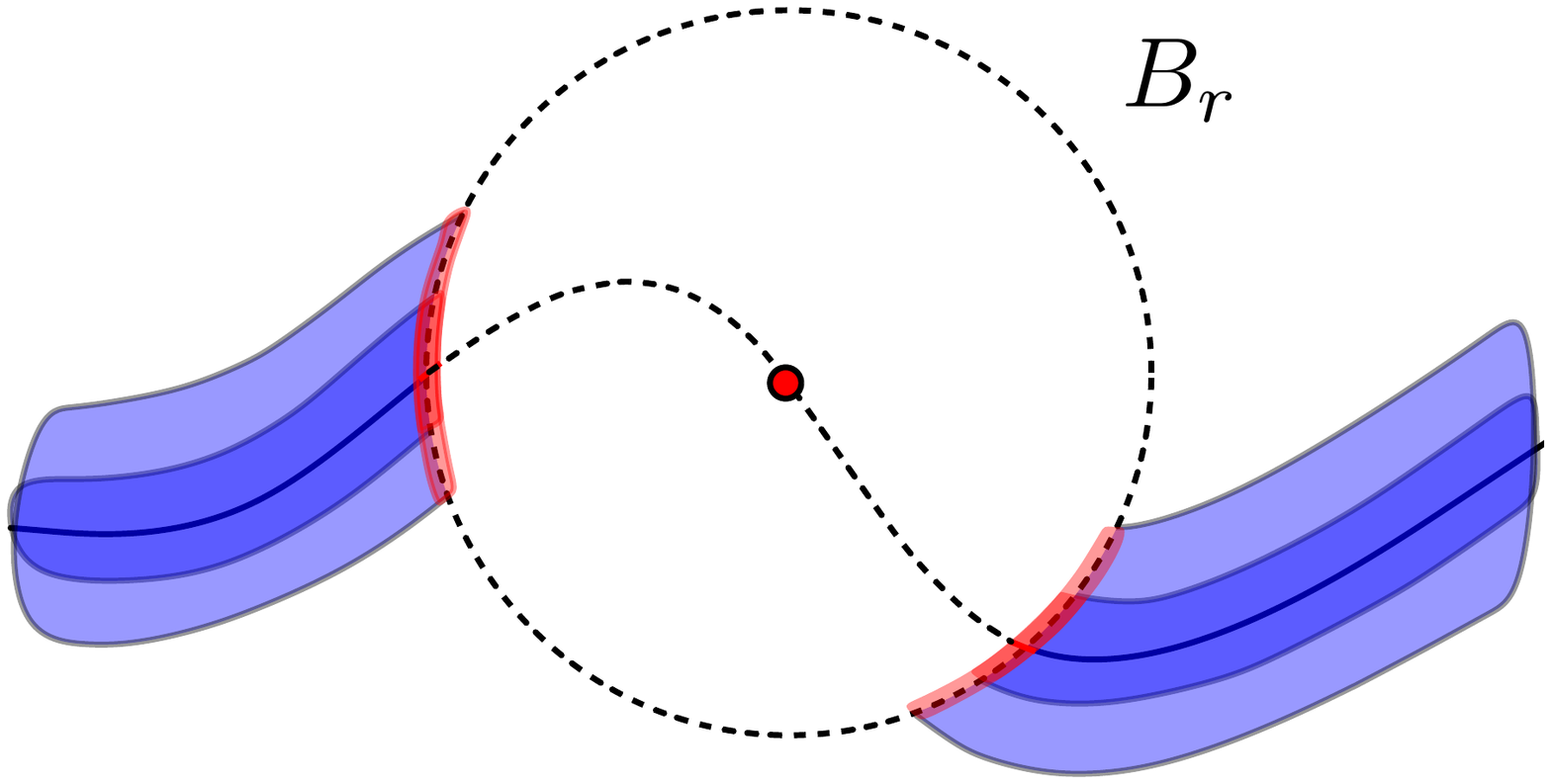} 
\caption{Left: the $\alpha$-filtration with respect to the pair $(\Xspace_{\alpha} \cap B_r, \Xspace_{\alpha} \cap \bdr B_r)$. Right: the filtration with respect to the pair 
$(\Xspace_{\alpha} , \Xspace_{\alpha} - \interior{B_r})$.}
\label{fig:filt_eq}
\end{center}
 \vspace{-0.5cm}
\end{figure}

We first show that the following two filtrations are equivalent (as $\alpha$ increases):
\begin{eqnarray}\label{eq:alphafilt1}
0 & \to & \Hgroup(\Xspace_{\alpha} \cap B_r, \Xspace_{\alpha} \cap \bdr B_r) \to \ldots \to \Hgroup(B_r, \bdr B_r), \\
0 & \to & \Hgroup(\Xspace_{\alpha} , \Xspace_{\alpha} -   \interior{B_r}) \to \ldots \to \Hgroup(\Rspace^n, \Rspace^n - \interior{B_r}). 
\label{eq:abs-b}
\end{eqnarray} 
Note that $\Xspace_{\alpha} - \interior{B_r} = \Xspace_{\alpha} - (\Xspace_{\alpha} \cap \interior{B_r})$. 
Unless otherwise specified, $\alpha \in [0,\infty)$. Graphically, these filtrations are shown in Fig.~\ref{fig:filt_eq}.  As it turns out, it is easier to argue about the filtration in Fig.~\ref{fig:filt_eq}(right) than Fig.~\ref{fig:filt_eq}(left), as shown in the following lemma. 
Recall a pair of space $(\Aspace, \Bspace)$ forms a \emph{good pair} if $\Bspace$ is a nonempty closed subspace that is a deformation retract of some neighborhood in $\Aspace$ (\cite{Hat2002}, page 114). 

\begin{lemma}
\label{lemma:excision}
Assuming that spaces $\Xspace_{\alpha} \cap B_r$ and  $ \Xspace_{\alpha} \cap
\bdr B_r$ form a \emph{good pair}, then $\Hgroup(\Xspace_{\alpha} \cap
B_r, \Xspace_{\alpha} \cap \bdr B_r) \cong
\Hgroup(\Xspace_{\alpha} , \Xspace_{\alpha} - \interior{B_r})$.
\end{lemma}
\proofsketch
This follows from the Excision Theorem  (\cite{GreHar1981}, Theorem 15.1, page 82) and the Excision Extension Theorem (\cite{GreHar1981}, Theorem 15.2, page 82). 
We excise the space $\Xspace_{\alpha} - B_r$ from the pair $(\Xspace_{\alpha} , \Xspace_{\alpha} - \interior{B_r})$, and obtain 
$\Hgroup(\Xspace_{\alpha}, \Xspace_{\alpha} - \interior{B_r}) \cong
\Hgroup(\Xspace_{\alpha} - (\Xspace_{\alpha} - B_r), \Xspace_{\alpha} - \interior{B_r} - (\Xspace_{\alpha} - B_r)) \cong \Hgroup(\Xspace_{\alpha} \cap
B_r, \Xspace_{\alpha} \cap \bdr B_r)$.
Since the closure of $\Xspace_{\alpha} - B_r$ needs not be contained in the interior of 
$\Xspace_{\alpha} - \interior{B_r}$, there are some technical conditions which require some care. 
See Appendix~\ref{sec-app:proofs} for details. 
\eop

We now show that we can approximate local homology at multi-scale 
via the $\alpha$-filtration using sample points. 
We begin with sequence (\ref{eq:abs-b}). Specifically, we first
consider the filtration corresponding to the whole space
$\{\Xspace_\alpha\}$,  and then the filtration corresponding to the
subspace we quotient by, $\{\Xspace_\alpha-\interior{B_r}\}$. The
key is a technical result described in Theorem \ref{theorem:quotient_interleave} 
which says that if we can approximate
filtrations independently, we can approximate their corresponding quotient filtration.
We consider this result to be of independent interest. 

\begin{theorem}
\label{theorem:quotient_interleave}
If we have two compatible filtrations interleaved with two other compatible 
filtrations, the relative filtration is also interleaved.
Formally, if compatible persistence modules $\Fcal = \{F_{\alpha}\}_{\alpha \in \Rspace}$ and $\Gcal= \{G_{\alpha}\}_{\alpha \in \Rspace}$ are
$\ep_1$-interleaved, $\Acal = \{A_{\alpha}\}_{\alpha \in \Rspace}$ and $\Bcal = \{B_{\alpha}\}_{\alpha \in \Rspace}$ are $\ep_2$-interleaved,
then the relative modules $\{(F_{\alpha},A_{\alpha})\}_{\alpha \in \Rspace}$ and $\{(G_{\alpha},B_{\alpha})\}_{\alpha \in \Rspace}$ are
$\ep$-interleaved, where $\ep = \max\{\ep_1, \ep_2\}$.
\end{theorem}
\proofsketch
Without loss of generality, assume $\ep_1 = \ep_2 = \ep$. 
Each pair, $\{(F,A)\}$ and $\{(G,B)\}$, gives rise to a long
exact sequence which are related by the interleaving maps. 
This gives the following commutative diagram:

\begin{center}
{\scalefont{0.9}
\begin{tikzcd}[column sep=small]
\Hgroup_n(A_\alpha) \arrow{r}{i^\alpha_n}  \arrow{d}{\phi^{\alpha}_n} & 
\Hgroup_n(F_\alpha) \arrow{r}{j_n^\alpha}  \arrow{d}{f_n^\alpha} & 
\Hgroup_n(F_\alpha,A_\alpha) \arrow{r}{k_n^{\alpha}}  \arrow{d}{\mu^\alpha_n} & 
\Hgroup_{n-1}(A_\alpha) \arrow{r}{i^{\alpha}_{n-1}}  \arrow{d}{\phi_{n-1}^{\alpha}} & 
{\Hgroup_{n-1}(F_{\alpha})}  \arrow{d}{f_{n-1}^\alpha} \\
\Hgroup_n(B_{\alpha+\ep}) \arrow{r}{p_n^{\alpha+\ep}}  \arrow{d}{\psi^{\alpha+\ep}_n} & \Hgroup_n(G_{\alpha+\ep}) \arrow{r}{q^{\alpha+\ep}_n}  \arrow{d}{g^{\alpha+\ep}_n} & \Hgroup_n(G_{\alpha+\ep},B_{\alpha+\ep}) \arrow{r}{r^{\alpha+\ep}_{n}}  \arrow{d}{\nu^{\alpha+\ep}_n} & 
\Hgroup_{n-1}(B_{\alpha+\ep}) \arrow{r}{p^{\alpha+\ep}_{n-1}}  \arrow{d}{\psi^{\alpha+\ep}_{n-1}} & 
\Hgroup_{n-1}(G_{\alpha+\ep}) \arrow{d}{g^{\alpha+\ep}_{n-1}}  \\
\Hgroup_n(A_{\alpha+2\ep}) \arrow{r}{i^{\alpha+2\ep}_n}  & 
\Hgroup_n(F_{\alpha+2\ep}) \arrow{r}{j^{\alpha+2\ep}_n}  & 
\Hgroup_n(F_{\alpha+2\ep},A_{\alpha+2\ep}) \arrow{r}{k^{\alpha+2\ep}_{n}} & 
\Hgroup_{n-1}(A_{\alpha+2\ep}) \arrow{r}{i^{\alpha+2\ep}_{n-1}}  & 
\Hgroup_{n-1}(F_{\alpha+2\ep})
\end{tikzcd}
}
\end{center} 
\vspace{-3mm}

To prove that the inter-leavings between individual modules imply 
an interleaving between $\{(F,A)\}$ and $\{(G,B)\}$, we would need some careful diagram chasing 
at the chain level. 
That is, we need to prove each of the four diagrams (reviewed in Fig.~\ref{fig:epinterleave}) needed for interleaving commutes,
i.e. diagrams in Fig. \ref{fig:lemma-eight-triangle} commute. 
The key issue is that
although each row is exact, maps between persistence modules do
not split --- therefore we may have one persistent relative class
without a persistent class in either component filtrations.  The
full details of the proof (with digram chasing arguments) are
given in Appendix~\ref{sec-app:proofs}.
\eop

\begin{figure}[!ht]
\vspace{-7mm}
\begin{center}
 \begin{tikzpicture}
\matrix[matrix of math nodes,column sep={50pt,between origins},row sep={30pt,between origins},nodes={anchor=center}] (s)
{
  |[name=rela]| \Hgroup_n(F_{\alpha},A_{\alpha}) &&|[name=relaa]| \Hgroup_n(F_{\alpha+2\epsilon},A_{\alpha+2\epsilon})& & |[name=rela1]| \Hgroup_n(F_{\alpha+\epsilon},A_{\alpha+\epsilon})&\\
&|[name=relb]| \Hgroup_n(G_{\alpha+\epsilon},B_{\alpha+\epsilon})&& |[name=relb1]| \Hgroup_n(G_{\alpha},B_{\alpha}) &&|[name=relbb1]| \Hgroup_n(G_{\alpha+2\epsilon},B_{\alpha+2\epsilon})\\
};
\draw[->] (rela) edge node[auto] {$$} (relaa)
          (relb1) edge node[auto] {$$} (relbb1)
          (rela) edge node[auto] {$$} (relb)
          (relb) edge node[auto] {$$} (relaa)
          (relb1) edge node[auto] {$$} (rela1)
          (rela1) edge node[auto] {$$} (relbb1);
\end{tikzpicture}
\vspace{-4mm}
\caption{Commuting diagrams for $\ep$-interleaved persistence modules.}
\label{fig:lemma-eight-triangle}
\end{center}
\vspace{-10mm}
\end{figure}

Before we state our main theorem below, we define 
the set of sample points which lie outside the ball, 
$\tilde{L} = \{p\in L | p\not\in B_r\}$, and $\tilde{L}_{\alpha} = \cup_{x \in \tilde{L}} B_{\alpha}(x)$.
We have:
\begin{theorem}
\label{theorem:alpha-sampling}
The persistence module with respect to the Vietoris-Rips filtration
of $\{(\Lspace_\alpha,\tilde\Lspace_\alpha)\}$, that is, 
$\{(\Rcal_{\alpha}(\Lspace),\Rcal_{\alpha}(\tilde\Lspace))\}$ is 
$\left(2\epsilon +\alpha + \frac{\alpha^2}{r}\right)$-interleaved with 
the $\alpha$-filtration, $\{(\Xspace_{\alpha}, \Xspace_{\alpha} - \interior B_r)\}$,  for $\alpha<r$.
\end{theorem}
\proofsketch
Since we would like to approximate the persistence
diagram of the pair $\{(\Xspace_{\alpha}, \Xspace_{\alpha} - \interior B_r)\}$, we could approximate each filtration
independently. 
We describe the key ingredients in our proof here and defer the technical details involving each step to Appendix \ref{sec-app:proofs}. 

\begin{figure}[!t]
\begin{center}
\vspace{-7mm}
\hspace{-10mm} 
\subfloat[]{\includegraphics[width=0.3\linewidth]{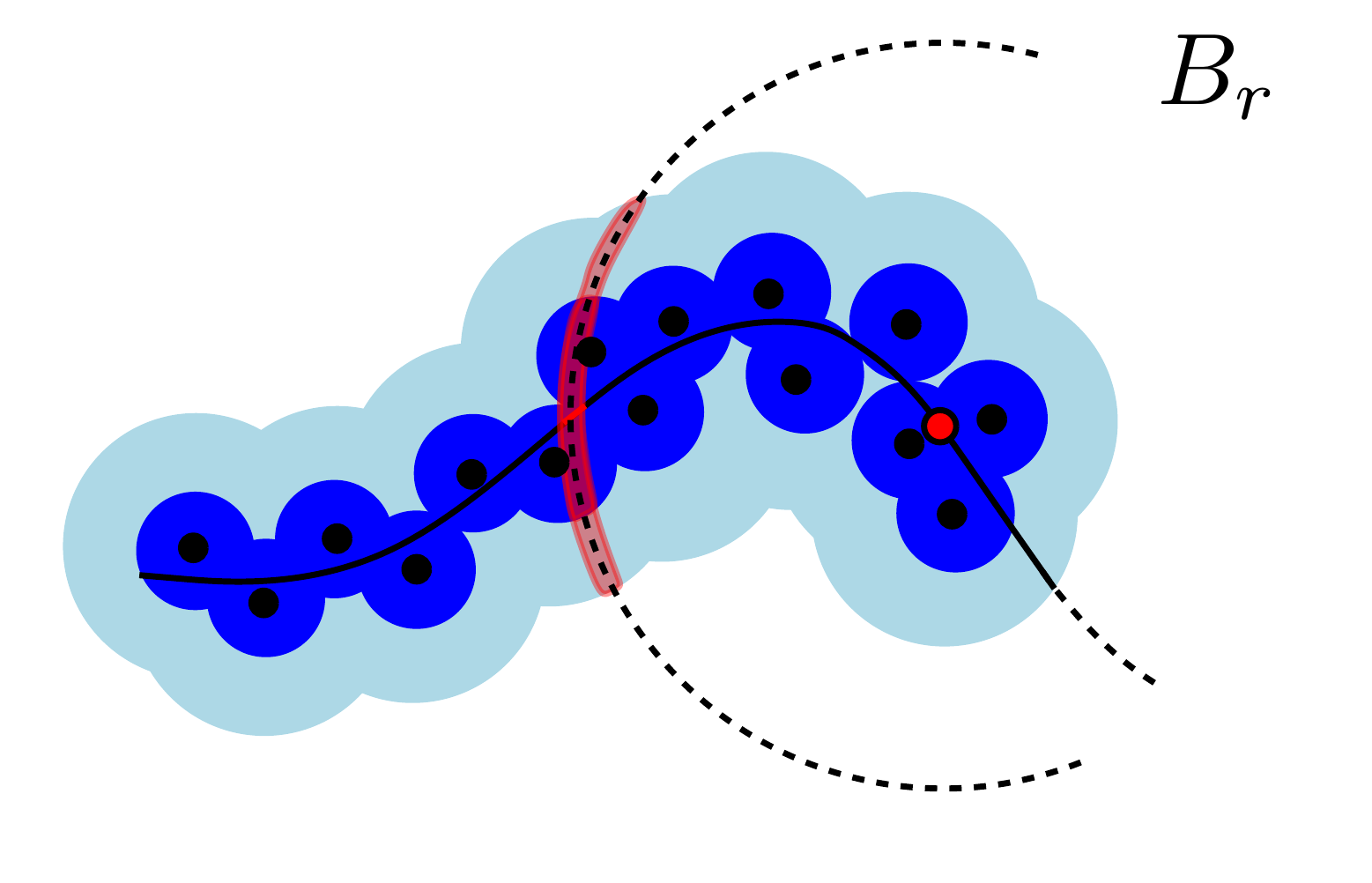}} \hspace{-6mm} 
\subfloat[]{\includegraphics[width=0.3\linewidth]{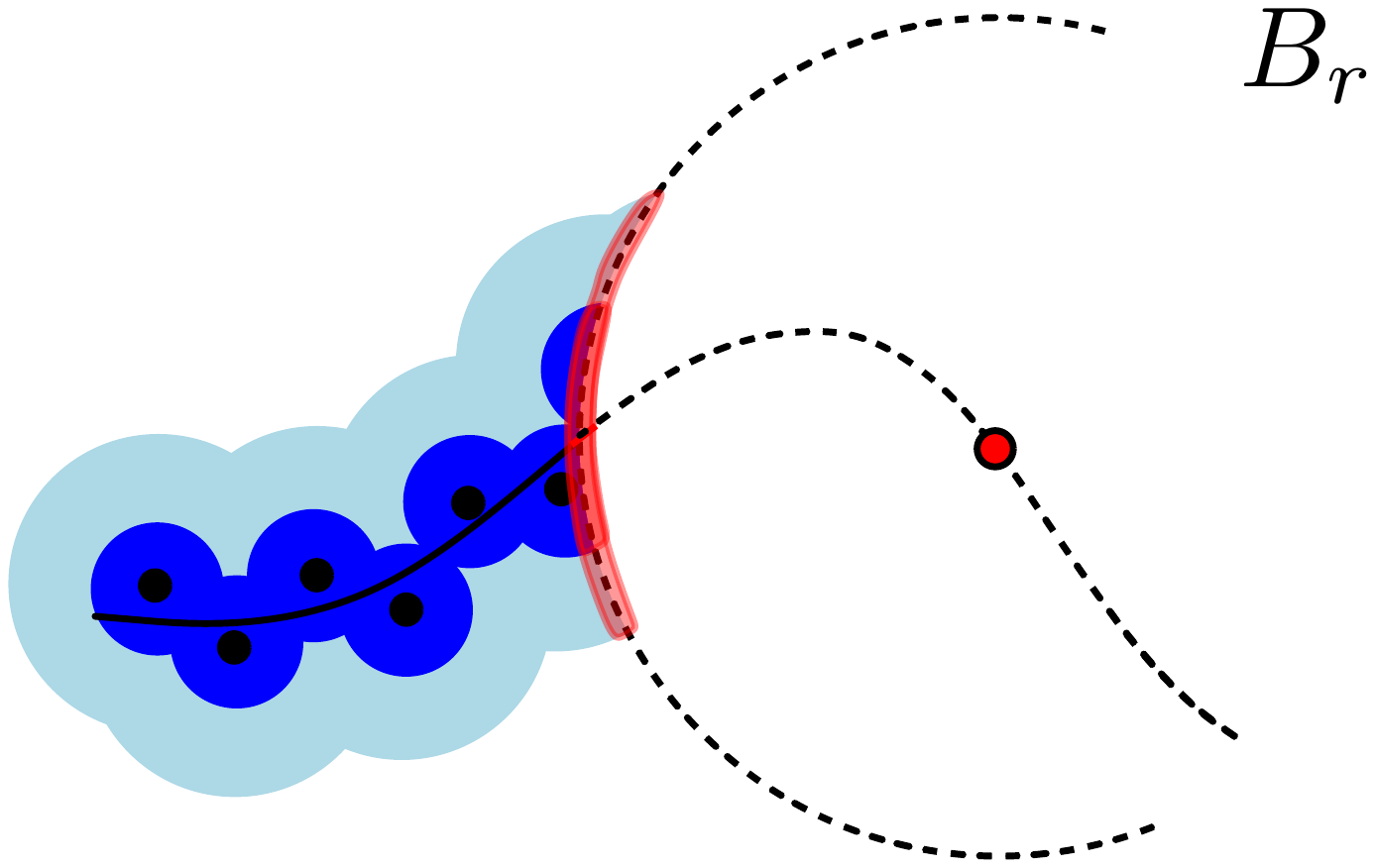}} \hspace{-6mm} 
\subfloat[]{\includegraphics[width=0.3\linewidth]{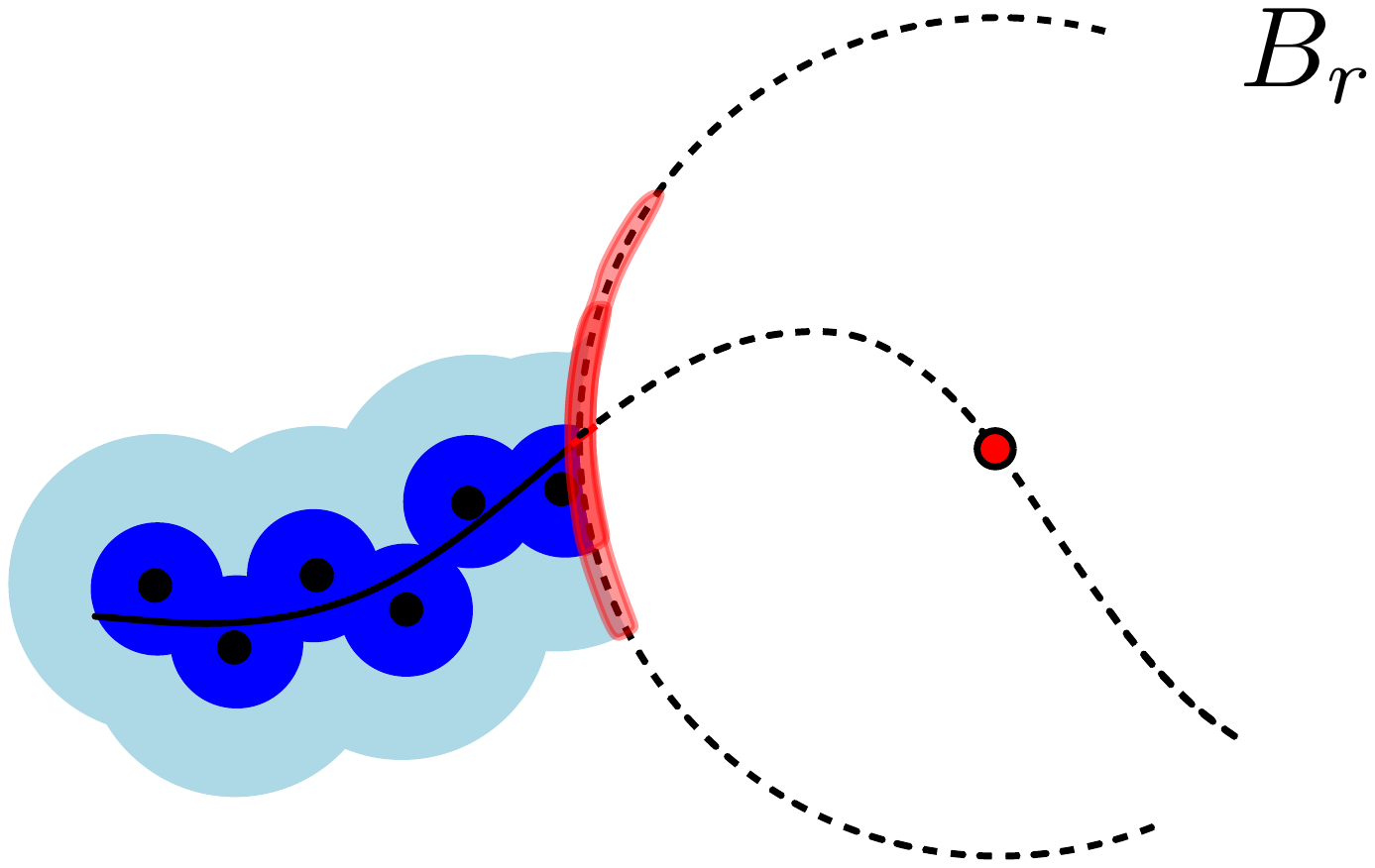}} \hspace{-6mm} 
\subfloat[]{\includegraphics[width=0.3\linewidth]{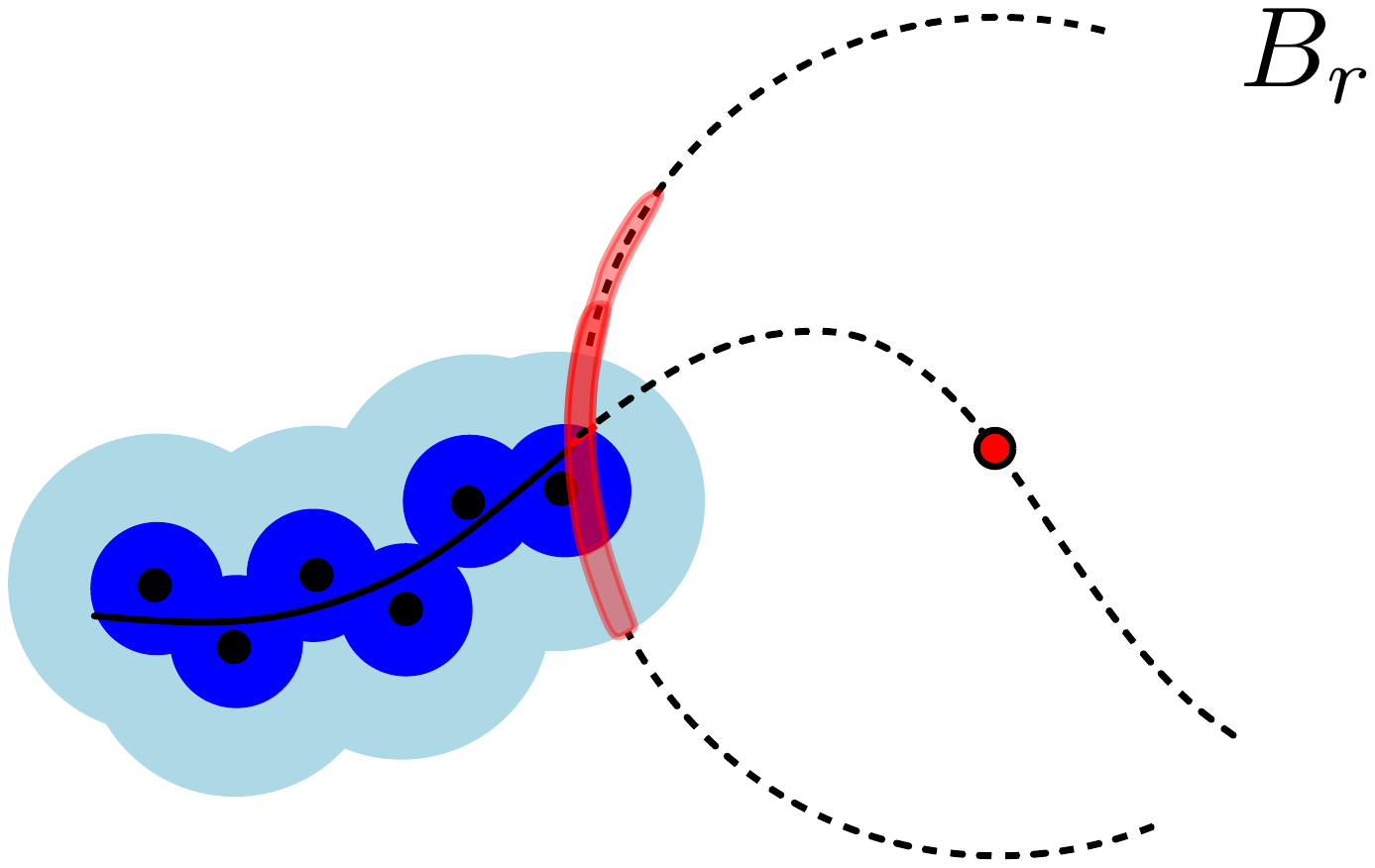}} \hspace{-6mm} 
\vspace{-3mm}
\caption{
(a) $L_\alpha$, (b) $L_{\alpha} - \interior B_r$,
(c) $\{\tilde\Lspace_{\alpha}\} - \interior B_r$,
(d) $\{\tilde\Lspace_{\alpha}\}$.
}
\label{fig:filt_proof}
\end{center}
 \vspace{-9mm}
\end{figure}

First, we consider the whole space filtration
$\{\Xspace_{\alpha}\}$, and show that $\{\Xspace_{\alpha}\}$ is
$\ep$-interleaved with $\{L_{\alpha}\}$ (Fig.~\ref{fig:filt_proof}(a)), which is relatively
straightforward assuming $L$ is an $\epsilon$-sample of $\Xspace$. 
Since the nerve of  $L_{\alpha}$ is the \v Cech complex $\Ccal_{\alpha}(L)$,
then  $\{\Xspace_{\alpha}\}$ is $\ep$-interleaved with $\{\Ccal_{\alpha}(L)\}$.
 
Second, approximating the subspace filtration
$\{\Xspace_\alpha-\interior{B_r}\}$ is more involved. The
straightforward approach is to simply remove $\interior{B_r}$
from $\{L_{\alpha}\}$ and consider 
$\{L_{\alpha} - \interior{B_r}\}$ (Fig.~\ref{fig:filt_proof}(b)). 
This is computational expensive, so instead, we consider 
$\{\tilde{L}_{\alpha} - \interior{B_r}\}$ (Fig.~\ref{fig:filt_proof}(c),  
note its subtle difference with Fig.~\ref{fig:filt_proof}(b)). 
We then show that $\{\Xspace_\alpha-\interior{B_r}\}$ is $2\ep$-interleaved
with $\{\tilde\Lspace_\alpha - \interior{B_r}\}$, by showing 
that $\tilde\Lspace$ is a $2\ep$-sample of $\Xspace -
\interior B_r$, that is, removing sampled points in the ball
gives a good sample of $\Xspace - \interior B_r$.  

Third, we further prove that $\{\tilde\Lspace_\alpha -\interior{B_r}\}$ is
$(\frac{\alpha^2}{r})$-interleaved with
$\{\tilde\Lspace_{\alpha}\}$. This is illustrated in
Fig.~\ref{fig:filt_proof}(d), where we allow the offset to
intersect inside the ball, and show that the error
remains controlled.  
In addition, the nerve of $\tilde\Lspace_{\alpha}$ is the \v Cech complex 
$\Ccal_{\alpha}(\tilde\Lspace)$. 
A combination of these results
implies that $\{\Xspace_\alpha-\interior{B_r}\}$ is $\left(2\ep
+\frac{\alpha^2}{r}\right)$-interleaved with $\{\Ccal_{\alpha}(\tilde \Lspace)\}$. 
%
Finally, having shown that we can approximate both the filtration
on the whole space and on the subspace (which we quotient by), we
invoke Theorem~\ref{theorem:quotient_interleave}. Finally, based
on that the Vietoris-Rips and \v{C}ech complexes are $\alpha$-interleaved, that is, 
$\Rcal_{\alpha} \subseteq \Ccal_{\alpha} \subseteq \Rcal_{2\alpha}$,
we obtain the additional $\alpha$ factor in the approximation
result. 
However, the limiting factor in this case is the subspace
filtration. The results are only meaningful over a range of
values for $\alpha$. For example, if $\alpha \geq r $, the local
homology is isomorphic to a $(d+1)$-sphere. This is discussed further in Appendix~\ref{sec-app:proofs}. 
\eop

\noindent\textbf{Computation.} Using Theorem~\ref{theorem:alpha-sampling}, we can compute relative persistent homology of
the filtrations built on the sample points using the algorithm
described in~\cite{SVJ2013}.




\vspace{-2mm}
\section{Approximating Local Homology: $r$-Filtration}
\vspace{-2mm}
\label{sec:r_filt}
In this section, we describe approximating local homology with respect to a fixed point $x$ at multi-scale via $r$-filtration (Fig. \ref{fig:r-filtration}).  We fix a thickening parameter $\alpha$ and drop it from
the notation, using only $\Xspace$. 
Consider the following filtration, for
$r \geq s \geq t$,
\begin{eqnarray}
\ldots \to  \Hgroup(\Xspace, \Xspace \cap B^r) \to  \Hgroup(\Xspace, \Xspace \cap B^s)  \to  \Hgroup(\Xspace, \Xspace \cap B^t) \to \ldots,
\label{eq:relrst}
\end{eqnarray}

Now we endow the space $\Xspace$ with a function $g: \Xspace
\to \Rspace$, which is the Euclidean distance to a fixed point $\p$,
$g(\p) = d(\p, y) = d_{\p}(y)$.  
$g$ could be viewed as the restriction onto the space $\Xspace$, of a Euclidean distance function 
 to a point $\p$, $d_x: \Rspace^d \to \Rspace$, that is, $g = \restr{d_x}{\Xspace}$.
The function $g$ is $1$-Lipschitz and 
we see that $\Xspace \cap B^r = g^{-1}[r, \infty)$, the
  superlevel set of $g$.  The above sequence becomes,
\begin{eqnarray}
\ldots \to \Hgroup(\Xspace, g^{-1}[r, \infty)) \to
  \Hgroup(\Xspace, g^{-1}[s, \infty)) \to \Hgroup(\Xspace,
    g^{-1}[t, \infty)) \ldots.
\label{eq:relrst2}
\end{eqnarray}
This is the relative persistence module of $g$.  Now let $f =
-g: \Xspace \to \Rspace$, $f$ is also $1$-Lipschitz.  
Sequence (\ref{eq:sublevel})
holds the same information as sequence (\ref{eq:relrst2}) assuming tame functions 
\footnote{It is unclear whether this holds in the case of non-tame
functions which could arise as a consequence of a pathological
underlying space.},
according to Extended Persistence Symmetry Corollary
\cite{CarlssonSilvaMorozov2009} (that is, the ordinary
persistence diagram of a function $f$ equals the relative
persistence diagram of $-f$ up to a dimension shift and central
reflection),
\begin{equation}
\ldots \to \Hgroup(f^{-1}(-\infty, a]) \to
    \Hgroup(f^{-1}(-\infty, b] ) \to \Hgroup(f^{-1}(-\infty, c] )
\ldots,
\label{eq:sublevel}
\end{equation}
where $a \leq b \leq c$, which corresponds to the persistence
module of $f$ based on its sublevel sets.
Since the filtrations in sequence (\ref{eq:sublevel}) and
sequence (\ref{eq:relrst2}) hold the same information, we can
translate the diagram and recover the information for the
original $r$-filtration (sequence (\ref{eq:relrst})).

The key insight is that in this case, the $r$-filtration amounts
to studying the persistent homology of a function on the space
--- the distance function to a point, which is a particularly nice
function, i.e. 1-Lipschitz.  
In this section, we give results under strong assumptions on the space $\Xspace$ with some further discussions deferred to Appendix \ref{sec-app:proofs-r}.  

We introduce a strong assumption on homotopy between a pair of
spaces, which requires that points are only moved a bounded
amount in the homotopy.  Two subsets of Euclidean space,
$\Xspace$ and $\Yspace$ are \emph{$\epsilon$-homotopy
  equivalent}, if there exists two functions
$i:\Xspace\to\Yspace$ and $h:\Yspace\to\Xspace$ such that 
$h\circ i$ is homotopic to the identity map $id_{\Xspace}$, 
$i\circ h$ is homotopic to $id_{\Yspace}$, 
$d(p,h\circ i(p) )\leq \ep$ and $d(p,i\circ h(p) )\leq \ep$.  
The consequence of such an assumption is discussed further in
Appendix \ref{sec-app:proofs-r}.
In our context, the map $i$ is typically the canonical inclusion map, 
therefore if $d(p,i\circ h(p) )\leq \ep$ then $d(p,h\circ i(p) )\leq \ep$. 
Then we refer to $h: \Yspace \to \Xspace$ as the \emph{$\ep$-homotopy equivalence} between 
$\Yspace$ and $\Xspace$, where $d(p, h(p)) \leq \ep$. 

The first step in approximating the $r$-filtration is relating
the sublevel set filtration of a $1$-Lipschitz function
$f: \Xspace \to \Rspace$ on the space $\Xspace$, and the sublevel
set filtration of a corresponding function
$f_{\ep}: \Xspace_{\ep} \to \Rspace$ on $\Xspace_\ep$.  
These filtrations together with maps induced by space inclusions form the (homology) persistence module of $f$ and $f_{\ep}$, respectively. 
Recall that $f$ is the negative of $d_x$ restricted to $\Xspace$, $f = - \restr{d_x}{\Xspace}$. 
Likewise, $f_{\ep} = -\restr{d_x}{\Xspace_{\ep}}$.  
Since there is an inclusion $\Xspace \hookrightarrow \Xspace_\ep$, it
follows that $f = \restr{f_{\ep}}{\Xspace}$.
For the rest of the section we use the following notation for sublevel sets:
$\F{a} = f^{-1}(-\infty,a]$, $\Fe{a} = f^{-1}_\ep(-\infty,a]$, for every 
$a \in \Rspace$.
The persistence module of $f$ and $f_{\ep}$ are represented as $\{\Hgroup(\F{a})\}_a$ 
and $\{\Hgroup(\Fe{a})\}_a$ respectively.

\begin{lemma}
\label{lem:offset_inter}
Suppose $\Xspace$ and $\Xspace_\ep$ are $\ep$-homotopy equivalent
through the canonical inclusion map $i: \Xspace \to \Xspace_\ep$ and the map $h: \Xspace_\ep \to \Xspace$. 
Then the persistence modules of $f$ and $f_{\ep}$, that is, 
$\{\Hgroup(\F{a})\}_a$ and $\{\Hgroup(\Fe{a})\}_a$, are $\ep$-interleaved. 
\end{lemma}
\proof
Consider the following sequence of maps: 
$$\F{\alpha} \xrightarrow{i'} \Fe{\alpha+\ep} \xrightarrow{h'} \F{\alpha+2\ep}.$$

We define the map $i' = \restr{i}{\F{\alpha}}$ and show that $i'$ is well-defined. 
$\forall p \in \F{\alpha}$, by definition, we have $f(p) \leq \alpha$ and 
$f = \restr{f_{\ep}}{\Xspace}$, therefore $f_{\ep}(p) \leq \alpha$. 
This implies that $p \in \Fe{\alpha} \subseteq \Fe{\alpha+\ep}$, therefore, 
$i'$ is a well-defined inclusion, which induces inclusion on the homology level, 
$i_*: \Hgroup(\F{\alpha}) \to \Hgroup(\Fe{\alpha+\ep})$. 

We define $h' = \restr{h}{\Fe{\alpha+\ep}}$, and we need to show that $h'$ is well-defined, that is, 
the image of $h'$ lies in $\F{\alpha+2\ep}$. 
$\forall p \in \Fe{\alpha+\ep}$, by definition, we have $f_{\ep}(p) \leq \alpha+\ep$. 
Since $f_{\ep} = - \restr{d_x}{\Xspace_{\ep}}$, then $- d(x,p) \leq \alpha+\ep$. 
Combining with $d(p, h(p)) \leq \ep$, we have 
$f(h(p)) := - d(x, h(p)) \leq d(p, h(p)) - d(x, p) \leq \alpha+2\ep$. 
This implies that $h(p) \in \F{\alpha+2\ep}$.
Therefore $h'$ is well-defined. 
In addition, based on our assumption that $\Xspace$ and $\Xspace_{\ep}$ are homotopy equivalent through maps $i$ and $h$, this implies that $h'$ is a homotopy equivalence, which induces an isomorphism $h_*$ on  the homology level, $h_*: \Hgroup(\Fe{\alpha+\ep}) \to \Hgroup(\F{\alpha+2\ep})$. 

In order to show persistence modules $\{\Hgroup(\F{a})\}_a$ 
and $\{\Hgroup(\Fe{a})\}_a$ are $\ep$-interleaved, we could easily verify that the four diagrams in Fig.~\ref{fig:epinterleave} commutes based on linear maps $i_*$ and $h_*$.  
\eop

The next step is to relate the above filtrations to the union of
balls on the samples. For notational convenience we define the
union of balls centered around points with a function value less
than some threshold  $a \in \Rspace$ as $\uball_\ep(a) = \cup_{p \in L, f(p) \leq a} B_\ep(p)$,
where $a \in \Rspace$ and $a \leq 0$. 
Since $\uball_\ep(a)$ contains Euclidean balls which are convex, the Never Lemma holds, 
that is, its nerve $\Ncal(\uball_\ep(a))$, which corresponds to the \v Cech complex $\Ccal_{\ep}(a)$, and $\uball_\ep(a)$ are homotopy equivalent.   
As $a$ varies, these complexes together with the maps induced by inclusions form a persistence module $\{\Hgroup(\Ccal_{\ep}(a))\}_{a}$.  
Similarly we define the corresponding Vietoris-Rips complex and its persistence module as $\Rcal_{\ep}(a)$ and $\{\Hgroup(\Rcal_{\ep}(a))\}_a$ respectively. 
  
\begin{lemma}
\label{lem:cech_inter}
Suppose $\Xspace$ and $\Xspace_\ep$ are $\ep$-homotopy equivalent
through the canonical inclusion map $i: \Xspace \to \Xspace_\ep$
and the map $h: \Xspace_\ep \to \Xspace$.  Suppose $\Lspace$ is
an $\ep$-sample of $\Xspace$.  Then the persistence modules $\{\Hgroup(\F{a})\}_a$ of
$f$ and $\{\Hgroup(\Ccal_{\ep}(a))\}_a$ are $2\ep$-interleaved.  
\end{lemma}

\proof 
The proof is nearly identical to the proof of
Lemma~\ref{lem:offset_inter}. Consider the following sequence: 
\begin{equation*}
F(\alpha) \xrightarrow{i'} \uball_\ep (\alpha+\ep) \xrightarrow{h'} F(\alpha+3\ep) 
\end{equation*}
We define the map $i' = \restr{i}{\F{\alpha}}$ and show $i'$ is well-defined. 
$\forall p \in \F{\alpha}$, by definition, $f(p) = -d(x,p) \leq \alpha$. 
Since $L$ is an $\ep$-sample of $\Xspace$, 
there exists $q \in L$ such that $ p \in B_{\ep}(q)$, 
that is, $d(p,q) \leq \ep$.
Combining the above inequalities, we obtain 
$f(q) = -d(x,q) \leq -d(x,p) + d(p,q) \leq \alpha+\ep$, 
implying that $p \in \uball_{\ep}(\alpha+\ep)$.

For map $h'$, since $\uball_\ep(\alpha+\ep)\subseteq F_\ep(\alpha+2\ep)$, based on the results in Lemma~\ref{lem:offset_inter} that the map $\Fe{\alpha+2\ep} \to \F{\alpha+3\ep}$ is well-defined,  we can
define $h' = h|_{\uball_\ep(\alpha+\ep)}$.
Following
similar argument in Lemma~\ref{lem:offset_inter}, $\{\Hgroup(\uball_\ep(a))\}_{a}$ is $2\ep$-interleaved
with $\{\Hgroup(\F{a})\}_{a}$. By the Nerve Lemma, the union
of balls\footnote{Note the definition of the union of balls filtration -- it precisely equals to the lower star filtration of the \v Cech complex.} is homotopic to the \v Cech complex for all $a$, 
leading to $\Hgroup(\Ccal_{\ep}(a)) \iso \Hgroup(\uball_\ep(a))$. 
\eop

\begin{theorem}
\label{theorem:r}
Suppose $h$ is an $2\ep$-homotopy equivalence between $\Xspace_{2\ep}$ and
$\Xspace$, and $\Lspace$ is an $\ep$-sample of $\Xspace$.
Then the Vietoris-Rips filtration $\{\Hgroup(\Rcal_{2\ep}(a))\}_a$ is a
$4\ep$-approximation of the $r$-filtration $\{\Hgroup(\F{a})\}_a$.
\end{theorem}
\proof
Suppose $\Xspace_{2\ep}$ and $\Xspace$ are $2\ep$-homotopy equivalent through the canonical inclusion map $i: \Xspace \to \Xspace_{2\ep}$ and the map $h: \Xspace_{2\ep} \to \Xspace$. 
We can construct the following commutative diagram on the space level:
\begin{center}
\resizebox{0.95\textwidth}{!}{
\begin{tikzpicture}[xscale=2,yscale=1]
 \node  (Xa) at (0,0.75) {$\F{\alpha}$};
 \node  (Ua) at (1,0) {$\uball_\ep(\alpha+\ep)$};
 \node (Ca) at (2,0.75) {$\Ccal_\ep(\alpha+\ep)$};
 \node (Ra) at (3,0.0) {$\Rcal_{2\ep}(\alpha+\ep)$};
 \node  (Ua1) at (5,0)  {$\uball_{2\ep}(\alpha+\ep)$};
 \node  (Ca1) at (4,0.75) {$\Ccal_{2\ep}(\alpha+\ep)$};
 \node  (Xa2) at (0,-1) {$\F{\alpha+5\ep}$};
 \node (Ua2) at (1,-1.75) {$\uball_\ep(\alpha+6\ep)$};
 \node  (Ca2) at (2,-1) {$\Ccal_\ep(\alpha+6\ep)$};
 \node (Ra2) at (3,-1.75) {$\Rcal_{2\ep}(\alpha+6\ep)$};
 \node  (Ua3) at (5,-1.75) {$\uball_{2\ep}(\alpha+6\ep)$};
 \node  (Ca3) at (4,-1) {$\Ccal_{2\ep}(\alpha+6\ep)$};
\draw[->] (Xa) -- (Ua);
\draw[->] (Ua) -- (Ca) node  [midway, above,xshift=-0.2cm,yshift=-0.15cm] {$\simeq$} ;
\draw[->] (Ca) -- (Ra);
\draw[->] (Ra) -- (Ca1);
\draw[<-] (Ca1) -- (Ua1) node  [midway, above,xshift=0.2cm,yshift=-0.15cm] {$\simeq$};;
\draw[->] (Xa2) -- (Ua2);
\draw[->] (Ua2) -- (Ca2) node [midway, above,xshift=-0.2cm,yshift=-0.15cm]  {$\simeq$} ;
\draw[->] (Ca2) -- (Ra2);
\draw[->] (Ra2) -- (Ca3);
\draw[<-] (Ca3) -- (Ua3) node [midway, above,xshift=0.2cm,yshift=-0.15cm] {$\simeq$};
\draw[->] (Xa) -- (Xa2);
\draw[->] (Ua) -- (Ua2);
\draw[->] (Ca) -- (Ca2);
\draw[->] (Ra) -- (Ra2);
\draw[->] (Ca1) -- (Ca3);
\draw[->] (Ua1) -- (Ua3);
\draw[->,gray,rounded corners] (Ua1) -- (5.5,0) -- (5.5,-0.5) -- (-0.75,-0.5) -- (-0.75,-1) -- (Xa2);
\end{tikzpicture}}
\end{center}
First we consider the top and bottoms rows in the diagram. 
The 1st map is an inclusion on the space level. 
The 2nd and 5th maps are homotopy equivalences based on the Nerve Lemma (which induces isomorphisms on the homology level). 
The 3rd and 4th maps are inclusions based on interleaving between \v Cech and Vietoris-Rips complexes, i.e. $\Ccal_{\ep} \subseteq \Rcal_{2\ep} \subseteq \Ccal_{2\ep}$. 
Second, all the vertical maps between the top and bottom rows are inclusions. 
Finally, we define the connecting map $\uball_{2\ep}(\alpha+\ep) \to \F{\alpha+5\ep}$ 
as $h' = \restr{h}{\uball_{2\ep}(\alpha+\ep)}$.
To show $h'$ is well-defined, $\forall p \in \uball_{2\ep}(\alpha+\ep)$, $f(p) \leq \alpha+3\ep$, 
since $h'$ is a $2\ep$-homotopy, $h'(p)$ has a function value at most $\alpha+5\ep$, 
therefore $h'(p) \in \F{\alpha+5\ep}$. 

From the above commutative diagram, we consider the following maps between spaces:
$\F{\alpha} \to \Rcal_{2\ep}(\alpha+\ep) \to \F{\alpha+5\ep}$.
This leads to a factor of $4\ep$ in the interleaving between persistence modules $\{\Hgroup(\Rcal_{2\ep}(a))\}_a$ and $\{\Hgroup(\F{a})\}_a$.  
\eop

\noindent\textbf{Computation.} Here we have reduced the computation of persistent local homology to standard persistence on the sample points \cite{EdeCohZom2002,ZomorodianCarlsson2005}.


\vspace{-3mm}
\section{Discussion}
\vspace{-2mm}

Local homology and relative homology are common tools in
algebraic topology.  In this paper, we recounted two different multi-scale
notions of local homology: the $\alpha$- and $r$-filtrations. 
We show that both can be well-approximated using Vietoris-Rips complexes based on a finite sample of the space and therefore  efficiently computed.  
We also prove a novel technical result involving interleaving between relative persistence modules 
derived from interleaving between absolute persistence modules. 
Several open questions remain: Are there better geometric
measures to describe the sampling conditions in approximating
local homology?  Could a similar sampling theory be developed for
witness complexes?  Under what conditions on the space are the
underlying filtrations we study tame? 

Our work was motivated by
stratification learning.  However the results in this paper could
be applied to any applications where local or relative homology
computations are relevant, i.e. for future directions, the
approximation of Conley index or well
groups \cite{ChazalSkrabaPatel2012}.


\bibliographystyle{splncs}
\bibliography{local-homology-refs}

\newpage
\appendix


\section{Detailed Proofs for Approximating Local Homology at Multi-scale via $\alpha$-filtration}
\label{sec-app:proofs}

\subsection{Lemma \ref{lemma:excision}}

\noindent\textbf{Lemma \ref{lemma:excision}.}
Assuming that spaces $\Xspace_{\alpha}$ and $ \Xspace_{\alpha} -
\interior{B_r}$ form a \emph{good pair},
then $\Hgroup(\Xspace_{\alpha} \cap B_r, \Xspace_{\alpha} \cap \bdr
B_r) \cong \Hgroup(\Xspace_{\alpha} , \Xspace_{\alpha} -
\interior{B_r})$.  
\proof 
First we recall several theorems
related to excisions.  Let $\Yspace, \Uspace, \Aspace$ be topological spaces.  The
inclusion map of pairs $(\Yspace - \Uspace, \Aspace - \Uspace ) \to (\Yspace, \Aspace) $ is called an
\emph{excision} if it induces a homology isomorphism.  In this
case, one says that $\Uspace$ can be excised.  We will make use of the
following two results about excision (\cite{GreHar1981}).

\begin{theorem}[Excision Theorem] (\cite{GreHar1981}, Theorem 15.1, page 82) 
If the closure of $\Uspace$ is contained in the interior of $\Aspace$, that is, $\closure{\Uspace} \subseteq \interior{\Aspace}$, then $\Uspace$ can be excised. 
\end{theorem}

\begin{theorem}[Excision Extension] (\cite{GreHar1981}, Theorem 15.2, page 82)
Suppose $\Vspace \subset \Uspace \subset \Aspace$ and 
(i) $\Vspace$ can be excised; (ii) $(\Yspace - \Uspace, \Aspace - \Uspace)$ is a deformation retract of $(\Yspace - \Vspace, \Aspace - \Vspace)$. Then $\Uspace$ can be excised. 
\label{thm:excext}
\end{theorem}

In our context, let $\Yspace = \Xspace_{\alpha}$, $\Aspace = \Xspace_{\alpha} -
\interior{B_r}$, $\Uspace = \Xspace_{\alpha} - B_r$.  
Therefore $\Yspace - \Uspace = \Xspace_{\alpha} \cap B_r$ and 
$\Aspace - \Uspace = \Xspace_{\alpha} \cap
\bdr B_r$.  However, since $\closure{\Uspace}$ needs not be contained in
$\interior{\Aspace}$, so we must define a suitable $\Vspace \subset \Uspace$.  One
direct way is to choose some small enough positive $\delta$ and a neighborhood $\Ispace$, 
such that we define, 
$\Ispace  =  \Xspace_{\alpha} \cap \bdr B_r \cap \closure{\Uspace}$, 
$\Ispace_{\delta}  =  \{x \in \closure{\Uspace} \mid d_{\Ispace}(x) \leq \delta \}$, 
and $\Vspace  =  \Uspace - \Ispace_\delta$,  
where $d_{\Ispace}(x)$ is the Euclidean distance from $x$ to the set $\Ispace$.

The existence of this $\delta$ follows from the assumption that the pair $(\Xspace_{\alpha}, \Xspace_{\alpha} - \interior{B_r}) := (\Yspace, \Aspace)$
form a \emph{good pair}. 
This is a technical condition which implies the existence of a neighborhood of 
$\Yspace - \Uspace$ (i.e. $\Yspace-\Vspace$) that deformation retracts to $\Yspace - \Uspace$. 
It is then straightforward to verify that $\Vspace \subset \Uspace \subset \Aspace$ satisfies the hypotheses of Theorem \ref{thm:excext}.

Therefore the chain map $k: C(\Yspace,\Aspace) \to C(\Yspace-\Uspace, \Aspace-\Uspace)$ is an excision. 
It is defined as $k = r_{\#} \circ s$, 
where $r_{\#}$ is the chain map induced by the retraction $r: (\Yspace-\Vspace, \Aspace-\Vspace) \to (\Yspace-\Uspace, \Aspace-\Uspace)$,
and $s$ is the chain-homotopy inverse of the chain map included by the inclusion of pairs 
$(\Yspace-\Vspace, \Aspace-\Vspace) \to (\Yspace,\Aspace)$, 
$s: C(\Yspace,\Aspace) \to C(\Yspace-\Vspace, \Aspace-\Vspace)$.
\eop

\subsection{Theorem \ref{theorem:quotient_interleave}}

We describe our long and technical proof of Theorem \ref{theorem:quotient_interleave} based on diagram chasing. 
We first need the following lemma that comes from the short exact sequences of a pair (\cite{Mun1984}, page 140). 
\begin{lemma}
\label{lemma:chain-pair}
The quotient map on the chain level commutes.  That is, for
compatible maps $A\to B$ and $X\to Y$ there is a map $(X, A) \to
(Y, B)$ such that the diagram in Fig. \ref{fig:short-exact-chain}
is commutative.
\end{lemma}

\begin{figure}[!ht]
\begin{center}
\begin{tikzpicture}
\matrix[matrix of math nodes,column sep={80pt,between origins},row sep={50pt,between origins},nodes={anchor=center}] (s)
{
|[name=zeroa]| 0  & |[name=A]| \Cgroup_n(A) & |[name=X]| \Cgroup_n(X) &|[name=rela]| \Cgroup_n(X,A)&|[name=zeroa1]| 0\\
|[name=zerob]| 0  & |[name=B]| \Cgroup_n(B) & |[name=Y]| \Cgroup_n(Y) &|[name=relb]| \Cgroup_n(Y,B )&|[name=zerob1]| 0\\
};
\draw[->] (zeroa) edge node[auto] {$$} (A)
          (A) edge node[auto] {$i$} (X)
          (X) edge node[auto] {$q$} (rela)
          (rela) edge node[auto] {$$} (zeroa1)
(zerob) edge node[auto] {$$} (B)
          (B) edge node[auto] {$j$} (Y)
          (Y) edge node[auto] {$r$} (relb)
          (relb) edge node[auto] {$$} (zerob1)
          (A) edge node[auto] {$f$} (B)
          (X) edge node[auto] {$g$} (Y)
           (rela) edge node[auto] {$h$} (relb)
;
\end{tikzpicture}
\caption{Commuting diagrams on the chain level.}
\label{fig:short-exact-chain}
\end{center}
\end{figure}

\proof The assumption of compatibility ensures the left square
commutes. Note that $i,j$ must be injective maps and in all the
case we consider $f$ and $g$ are also injective, which is
sufficient for compatibility. To define $h$ we note that $ \image
h=\image g / (\image (g \circ i) \oplus \image j) $.  To show
that the right square commutes ($h\circ q = r\circ g$), we note
that any class in $\image(r\circ g)$ must be in $\image q$ by
exactness and the assumption that the left square commutes ($g
\circ i = j\circ f$). Since it is not in $\image i$ or map to
$\image j$, it is in $\image h$. Alternatively, any class in
$\image (h\circ q)$ must have a lift to $\Cgroup(Y)$ since $r$ is
a surjection. This must be in $\image g$ by the definition of
$h$, which concludes the proof.
  \eop

\noindent\textbf{Theorem \ref{theorem:quotient_interleave}.}
If we have two compatible filtrations interleaved with two other compatible 
filtrations, the relative filtration is also interleaved.
Formally, if compatible persistence modules $\Fcal = \{F_{\alpha}\}_{\alpha \in \Rspace}$ and $\Gcal= \{G_{\alpha}\}_{\alpha \in \Rspace}$ are
$\ep_1$-interleaved, $\Acal = \{A_{\alpha}\}_{\alpha \in \Rspace}$ and $\Bcal = \{B_{\alpha}\}_{\alpha \in \Rspace}$ are $\ep_2$-interleaved,
then the relative modules $\{(F_{\alpha},A_{\alpha})\}_{\alpha \in \Rspace}$ and $\{(G_{\alpha},B_{\alpha})\}_{\alpha \in \Rspace}$ are
$\ep$-interleaved, where $\ep = \max\{\ep_1, \ep_2\}$.

\proof
We begin with a list of notations.   
Suppose $\{F\}$ and $\{G\}$ are compatible and are $\ep$-interleaved with homomorphisms 
$\{f_{\alpha}: \Hgroup(F_{\alpha}) \to \Hgroup(G_{\alpha+\ep})\}$ and 
$\{g_{\alpha}: \Hgroup(G_{\alpha}) \to \Hgroup(F_{\alpha+\ep})\}$.  
Suppose $\{A\}$ and $\{B\}$ are also compatible and $\ep$-interleaved, with homomorphisms $
\{\phi_{\alpha}: \Hgroup(A_{\alpha}) \to \Hgroup(B_{\alpha+\ep})\}$ and 
$\{\psi_{\alpha}: \Hgroup(B_{\alpha}) \to \Hgroup(A_{\alpha+\ep})\}$.  
For relative homology to be well-defined,
we have injective maps at chain level, for simplicity, we further require 
$A_{\alpha} \hookrightarrow F_{\alpha}$ and 
$B_{\alpha} \hookrightarrow G_{\alpha}$.

We would like to prove that 
$\{(F, A)\}$ and $\{(G,B)\}$ are also interleaved, and we
could construct their corresponding homomorphisms,
$\{\mu_{\alpha}: \Hgroup(F_{\alpha}, A_{\alpha}) \to \Hgroup(F_{\alpha+\ep}, A_{\alpha+\ep})\}$ and
$\{\nu_{\alpha}: \Hgroup(G_{\alpha}, B_{\alpha}) \to \Hgroup(G_{\alpha+\ep}, B_{\alpha+\ep})\}$.

\begin{figure}[!ht]
\begin{center}
{\scalefont{1.0}
\begin{tikzcd}
\Hgroup_n(A_\alpha) \arrow{r}{i^\alpha_n}  \arrow{d}{\phi^{\alpha}_n} & 
\Hgroup_n(F_\alpha) \arrow{r}{j_n^\alpha}  \arrow{d}{f_n^\alpha} & 
\Hgroup_n(F_\alpha,A_\alpha) \arrow{r}{k_n^{\alpha}}  \arrow{d}{\mu^\alpha_n} & 
\Hgroup_{n-1}(A_\alpha) \arrow{r}{i^{\alpha}_{n-1}}  \arrow{d}{\phi_{n-1}^{\alpha}} & 
{\Hgroup_{n-1}(F_{\alpha})}  \arrow{d}{f_{n-1}^\alpha} \\
\Hgroup_n(B_{\alpha+\ep}) \arrow{r}{p_n^{\alpha+\ep}}  \arrow{d}{\psi^{\alpha+\ep}_n} & \Hgroup_n(G_{\alpha+\ep}) \arrow{r}{q^{\alpha+\ep}_n}  \arrow{d}{g^{\alpha+\ep}_n} & \Hgroup_n(G_{\alpha+\ep},B_{\alpha+\ep}) \arrow{r}{r^{\alpha+\ep}_{n}}  \arrow{d}{\nu^{\alpha+\ep}_n} & 
\Hgroup_{n-1}(B_{\alpha+\ep}) \arrow{r}{p^{\alpha+\ep}_{n-1}}  \arrow{d}{\psi^{\alpha+\ep}_{n-1}} & 
\Hgroup_{n-1}(G_{\alpha+\ep}) \arrow{d}{g^{\alpha+\ep}_{n-1}}  \\
\Hgroup_n(A_{\alpha+2\ep}) \arrow{r}{i^{\alpha+2\ep}_n}  & 
\Hgroup_n(F_{\alpha+2\ep}) \arrow{r}{j^{\alpha+2\ep}_n}  & 
\Hgroup_n(F_{\alpha+2\ep},A_{\alpha+2\ep}) \arrow{r}{k^{\alpha+2\ep}_{n}} & 
\Hgroup_{n-1}(A_{\alpha+2\ep}) \arrow{r}{i^{\alpha+2\ep}_{n-1}}  & 
\Hgroup_{n-1}(F_{\alpha+2\ep})
\end{tikzcd}
}
\end{center}
\caption{Commuting diagrams for the long exact sequence involving two pairs of filtrations.}
\label{fig:lemma-eight}
\end{figure}

To prove the result, we pass to the stack of long exact sequences in Fig. \ref{fig:lemma-eight}. 
First, we explain  the notation. 
A map, i.e. $\phi_n^{\alpha}$, represents a map that maps 
$n$-dimensional homology groups of $A_{\alpha}$ to some other homology groups.  
We note that all the squares in this diagram (Fig. \ref{fig:lemma-eight}) commute based on Lemma \ref{lemma:chain-pair}, 
and by
assumption the two component filtrations are interleaved, so the
first, second, fourth and fifth columns commute with the maps
induced by inclusion. 
For example, the map induced by inclusion $\image(\Hgroup_n(F_\alpha)\rightarrow
\Hgroup_n(F_{\alpha+2\epsilon}))$ equals $\image(g_{n}^{\alpha+\epsilon}
\comp f_n^{\alpha})$. 
Commutativity implies interleaving in some of the cases. 
We prove the following triangle commutes (Fig. \ref{fig:lemma-eight-triangle}) through four claims. 


\textbf{Claim 1}: if a relative class is in
$\image(\Hgroup_n(F_\alpha,A_\alpha)\rightarrow
\Hgroup_n(F_{\alpha+2\epsilon},A_{\alpha+2\epsilon}))$, and it  is in
$\image j_n^\alpha$ and $\image j_n^{\alpha+2\epsilon}$, then 
it is in $\image q_n^{\alpha+\ep}$.

If a relative class $\gamma$ in
$\image(\Hgroup_n(F_\alpha,A_\alpha)\rightarrow
\Hgroup_n(F_{\alpha+2\epsilon},A_{\alpha+2\epsilon}))$  is in
$\image j_n^\alpha$ and $\image j_n^{\alpha+2\epsilon}$, then by
the interleaving, it must be in
$\Hgroup_n(G_{\alpha+\epsilon})$. 
Therefore suppose $\gamma$ is not in
$\image q_n^{\alpha+\epsilon}$, it must have a preimage in
$\Hgroup_n(B_{\alpha+\epsilon})$. 
Since $\gamma$ is in $\image j_n^{\alpha+2\epsilon}$, 
it does not have a preimage in  $\Hgroup_n(A_{\alpha+2\epsilon})$. 
This would imply that
the lower left square does not commute ($
g_{n}^{\alpha+\epsilon}\circ p_{n}^{\alpha+\epsilon} \neq
i_{n}^{\alpha+2\epsilon} \circ \psi_{n}^{\alpha+\epsilon} $).
That is a contradiction, therefore it must be in $\image q_n^{\alpha+\ep}$. 

\textbf{Claim 2}: If a relative class is in $\image(\Hgroup_n(F_\alpha,A_\alpha)\rightarrow
\Hgroup_n(F_{\alpha+2\epsilon},A_{\alpha+2\epsilon}))$, and it is in
$\cokernel j_n^\alpha$ and $\cokernel j_n^{\alpha+2\epsilon}$, 
it must be in $\cokernel q_n^{\alpha+\ep}$.
%
If the relative class $\gamma$ in $\image(\Hgroup_n(F_\alpha,A_\alpha)\rightarrow
\Hgroup_n(F_{\alpha+2\epsilon},F_{\alpha+2\epsilon}))$ is in
$\cokernel j_n^\alpha$ and $\cokernel j_n^{\alpha+2\epsilon}$, then by
exactness $\gamma$ maps into  $\image k_n^{\alpha}$ and $\image k_n^{\alpha+2\ep}$,
that is, it maps to a non-trivial element in $\Hgroup_{n-1}(A_{\alpha})$
and $\Hgroup_{n-1}(A_{\alpha+2\epsilon})$. 
By the interleaving between $A$ and $B$, it must also map to an element of
$\Hgroup_{n-1}(B_{\alpha+\epsilon})$. 
Furthermore, it must be in $\kernel i_{n-1}^{\alpha}$.
Therefore suppose $\gamma$ is not in $\cokernel q_n^{\alpha+\ep}$
(or equivalently, $\image r_n^{\alpha+\epsilon}$ or $\kernel p_{n-1}^{\alpha+\ep}$) , it must map to a class in
$\Hgroup_{n-1}(G_{\alpha+\epsilon})$, which implies that the top
right square does not commute ($f_{n-1}^\alpha \circ
i_{n-1}^\alpha \neq p_{n-1}^{\alpha+\epsilon} \circ
\phi_{n-1}^{\alpha+\epsilon} $) leading to a contradiction.

We now show that commutativity is not a sufficient argument. Consider a
persistent relative class in $\Hgroup_n(F_\alpha,A_\alpha)\rightarrow
\Hgroup_n(F_{\alpha+2\epsilon},A_{\alpha+2\epsilon})$ such that it is
in $\image j_n^\alpha$ and 
$\cokernel j_n^{\alpha+2\epsilon}$. Alternatively, it may be in 
$\cokernel j_n^\alpha$ and $\image j_n^{\alpha+2\epsilon}$. In these cases, we
may map this class to zero the middle row and still maintain the
commutativity of the diagram (although this implies the relative
filtrations are not interleaved). This problem stems from the
fact that the maps between persistent modules do not \emph{split}
(The relative persistence module does not split into direct sum
of the image and cokernel in the long exact sequence).

\textbf{Claim 3}: If the relative class is in $\image(\Hgroup_n(F_\alpha,A_\alpha)\rightarrow
\Hgroup_n(F_{\alpha+2\epsilon},A  _{\alpha+2\epsilon}))$, then it is not possible that it is in
$\image j_n^{\alpha}$ and $\cokernel j_n^{\alpha+2\epsilon}$ at the same time.

First we handle the case where the relative class is in 
$\image j_n^\alpha$ and $\cokernel j_n^{\alpha+2\epsilon}$ by showing this
cannot occur. Since it is in $\image j_n^\alpha$ at the chain level,
there is a cycle representative in $\Zgroup_n(F_{\alpha})$. Since this
maps to a cycle representative in $\Zgroup_n(F_{\alpha+2\epsilon})$,
this implies that the cycle is in the boundary. However, looking
at the relavent part of the short exact sequence shown in Fig. \ref{fig:short-exact-chain2}.


\begin{figure}
\begin{center}
 \begin{tikzpicture}
\matrix[matrix of math nodes,column sep={40pt},row sep={40pt,between origins},nodes={anchor=center}] (s)
{
|[name=Fa]| \Cgroup_{n+1}(F_{\alpha+2\alpha}) &|[name=rela]| \Cgroup_{n+1}(F_{\alpha+2\epsilon},A_{\alpha+2\epsilon})\\
|[name=Fa1]| \Cgroup_{n}(F_{\alpha+2\alpha}) &|[name=rela1]| \Cgroup_{n}(F_{\alpha+2\epsilon},A_{\alpha+2\epsilon})\\
};
\draw[->] (Fa) edge node[auto] {$q$} (rela)
          (Fa1) edge node[auto] {$q$} (rela1)
(Fa) edge node[auto] {$\partial$} (Fa1)
          (rela) edge node[auto] {$\partial$} (rela1)
;
\end{tikzpicture}\caption{Short exact sequence on chain level.}
\label{fig:short-exact-chain2}
\end{center}
\end{figure}

The cycle representative in $\Cgroup_{n}(F_{\alpha+2\epsilon})$ lifts
to some element in $\Cgroup_{n+1}(F_{\alpha+2\epsilon})$. Now by
assumption, there is still some cycle representative in
$\Cgroup_{n}(F_{\alpha+2\epsilon},A_{\alpha+2\epsilon})$. By
commutativity, the bounding element in
$\Cgroup_{n+1}(F_{\alpha+2\epsilon})$ must also map to a bounding
element of the cycle representative in
$\Cgroup_{n}(F_{\alpha+2\epsilon},A_{\alpha+2\epsilon} )$, meaning it
cannot be a relative homology class. If on the other hand, the
cycle representative in $\Cgroup_{n}(F_{\alpha+2\epsilon})$ is in the
kernel of the quotient map, a relative homology class would
appear one dimension up. This is the case we deal with next.

\textbf{Claim 4}: If the relative class is in $\image(\Hgroup_n(F_\alpha,A_\alpha)\rightarrow
\Hgroup_n(F_{\alpha+2\epsilon},F_{\alpha+2\epsilon}))$, and it is in $\cokernel j_n^\alpha$ and 
$\image j_n^{\alpha+2\epsilon}$, then it must be in $\image{q_n^{\alpha+\ep}}$ or  $\cokernel{q_n^{\alpha+\ep}}$ (i.e. it must be $\Hgroup_n(G_{\alpha+\ep},B_{\alpha+\ep})$). 

For a relative class in $\cokernel j_n^\alpha$, there is a cycle
representative in $\Cgroup_{n-1}(A_\alpha)$ of the corresponding
class $\Hgroup_{n-1}(A_\alpha)$ which by the injectivity of the
interleaving, maps to a cycle in $\Cgroup_{n-1}(B_{\alpha+\ep})$
and $\Cgroup_{n-1}(A_{\alpha+2\ep})$. Further, since it is in
$\cokernel j_n^\alpha$, it follows that it maps to a bounded
cycle in $\Cgroup_{n-1}(F_\alpha)$ (and by injectivity) the
corresponding cycle representatives in
$\Cgroup_{n-1}(G_{\alpha+\ep})$ and
$\Cgroup_{n-1}(F_{\alpha+2\ep})$ are also bounded. Since this
relative class is assumed to be in $\image j_n^{alpha+2\ep}$, it
follows that the cycle representative in
$\Cgroup_{n-1}(A_{\alpha+2\ep})$ is now bounded, with the
pre-boundary mapping to a cycle in
$\Cgroup_{n}(F_{\alpha+2\ep})$. This follows from a chain level
understanding of the exactness of the bottom row.  Take the
representative $(n-1)$-cycle in $\Cgroup_{n-1}(A_{\alpha})$
denoted by $a$ and map it into
$\Cgroup_{n-1}(F_{\alpha})$. $i^\alpha_{n-1}(a)$ has a
pre-boundary in $\Cgroup_{n}(F_{\alpha})$ which maps to the cycle
representative of the relative class in $\Cgroup_{n}(F_{\alpha},
A_\alpha)$. This is just the connecting homomorphism
construction. If we map this relative cycle representative to
$\Cgroup_{n}(F_{\alpha+2\ep}, A_{\alpha+2\ep})$, since the class
is in $j_n^{\alpha+2\ep}$, it lifts to a non-trivial cycle in
$\Cgroup_{n}(F_{\alpha+2\ep})$. This cycle is precisely the image
of the pre-boundary of $a$ in $\Cgroup_{n-1}(A_{\alpha+2\ep})$
mapped to $\Cgroup_{n}(F_{\alpha+2\ep})$ plus the pre-boundary of the image
of $i^{\alpha}_{n-1}(a)$ in $\Cgroup_{n-1}(F_{\alpha+2\ep})$.

There are two case to consider. If $\phi^{\alpha}_{n-1}(a)$ is a
non-trivial cycle, then there is a homology class in $\kernel
p_{n-1}^{alpha+\ep}$ and by exactness, a corresponding class in
the $\cokernel q^{\alpha+\ep}_n$.

If $\phi^{\alpha}_{n-1}(a)$ maps to a bounded cycle, then by the
same reasoning as above, the pre-boundary of this cycle in
$\Cgroup_n^(B^{\alpha+\ep}_n$ must map to a non-trivial cycle in
$\Cgroup_n^(G^{\alpha+\ep}_n$. Hence there is a corresponding
class in $\image q^{\alpha+\ep}_n$. Proving the claim. 


Following the above four claims, we've shown the triangle in Fig. \ref{fig:lemma-eight-triangle} commutes. 
Fig.  \ref{fig:lemma-eight-triangle} equals the trapezoid in Fig.~\ref{fig:lemma-eight-2}(a) by setting $\alpha' = \alpha$. 
It follows that the trapezoid in Fig.~\ref{fig:lemma-eight-2}(a) commutes based on similar diagram chasing argument.

The other diagrams in Fig. \ref{fig:lemma-eight-2} follow similar proofs. 
For example, to show that the diagram in Fig.~\ref{fig:lemma-eight-2}(d) commutes, 
the argument goes through in precisely the same way, on diagrams  shown in Fig. \ref{fig:lemma-eight-3} 
and Fig. \ref{fig:lemma-eight-4}. 

\begin{figure}[!ht]
\begin{center}
\begin{tikzpicture}
\matrix[matrix of math nodes,column sep={0pt},row sep={30pt,between origins},nodes={anchor=center}] (s)
{
|[name=trapa]| \Hgroup(F_{\alpha},A_{\alpha)}  && & &|[name=trapb]|  \Hgroup(F_{\alpha'+2\epsilon},A_{\alpha'+2\epsilon}) \\
&|[name=trapc]| \Hgroup(G_{\alpha+\epsilon},B_{\alpha+\epsilon})  & &|[name=trapd]| \Hgroup(G_{\alpha'+\epsilon},B_{\alpha'+\epsilon}) &\\
&&(a)&&\\
&|[name=trapc1]|  \Hgroup(F_{\alpha+\epsilon},A_{\alpha+\epsilon}) & &|[name=trapd1]|   \Hgroup(F_{\alpha+\epsilon},A_{\alpha'+\epsilon})\\
|[name=trapa1]| \Hgroup(G_{\alpha},B_{\alpha})  & && &|[name=trapb1]|  \Hgroup(G_{\alpha'+2\epsilon},B_{\alpha'+2\epsilon})  &\\
&&(b)&&\\
& |[name=onea]| \Hgroup(F_{\alpha+\epsilon},A_{\alpha'+\epsilon})  & &&|[name=oneb]|  \Hgroup_n(F_{\alpha'+\epsilon},A_{\alpha'+\epsilon})\\
  |[name=onec]|  \Hgroup(G_{\alpha},B_{\alpha})  & &&|[name=oned]|   \Hgroup(G_{\alpha'},B_{\alpha'})\\
&&(c)&&\\
 |[name=onec1]|  \Hgroup(F_{\alpha},A_{\alpha})  & &&|[name=oned1]|   \Hgroup(F_{\alpha'},A_{\alpha'})&\\
& |[name=onea1]| \Hgroup(G_{\alpha+\epsilon},B_{\alpha+\epsilon}) & & &|[name=oneb1]|  \Hgroup(G_{\alpha'+\epsilon},B_{\alpha'+\epsilon})\\
&&(d)&&\\
};
\draw[->] (trapa) edge node[auto] {$$} (trapb)
          (trapc) edge node[auto] {$$} (trapd)
          (trapa) edge node[auto] {$$} (trapc)
          (trapd) edge node[auto] {$$} (trapb)
         (trapa1) edge node[auto] {$$} (trapb1)
          (trapc1) edge node[auto] {$$} (trapd1)
          (trapa1) edge node[auto] {$$} (trapc1)
          (trapd1) edge node[auto] {$$} (trapb1)
          (onea) edge node[auto] {$$} (oneb)
          (onec) edge node[auto] {$$} (oned)
          (onec) edge node[auto] {$$} (onea)
          (oned) edge node[auto] {$$} (oneb)
         (onea1) edge node[auto] {$$} (oneb1)
          (onec1) edge node[auto] {$$} (oned1)
          (onec1) edge node[auto] {$$} (onea1)
          (oned1) edge node[auto] {$$} (oneb1)
;
\end{tikzpicture}
\caption{Commuting diagrams for $\ep$-leaving of the pairs.}
\label{fig:lemma-eight-2}
\end{center}
 \vspace{-0.5cm}
\end{figure}

\begin{figure}[!ht]
\begin{center}
\begin{tikzpicture}
\matrix[matrix of math nodes,column sep={30pt},row sep={40pt,between origins},nodes={anchor=center}] (s)
{
|[name=Aa]| \Hgroup(A_{\alpha})  & |[name=Fa]| \Hgroup(F_{\alpha})  &|[name=rela]| \Hgroup(F_{\alpha},A_{\alpha})  & |[name=Aa1]| \Hgroup(A_{\alpha})& |[name=Fa1]| \Hgroup(F_{\alpha})\\
|[name=Ba]| \Hgroup(A_{\alpha'})  & |[name=Ga]| \Hgroup(F_{\alpha'})  &|[name=relb]| \Hgroup(F_{\alpha'},A_{\alpha'})  & |[name=Ba1]| \Hgroup(A_{\alpha'})& |[name=Ga1]| \Hgroup(F_{\alpha'})\\
|[name=Aaa]| \Hgroup(B_{\alpha'+\epsilon})  & |[name=Faa]| \Hgroup(G_{\alpha'+\epsilon})  &|[name=relaa]| \Hgroup(G_{\alpha'+\epsilon},B_{\alpha'+\epsilon})  & |[name=Aaa1]| \Hgroup(B_{\alpha'+\epsilon})& |[name=Faa1]| \Hgroup(G_{\alpha'+\epsilon})\\
};
\draw[->] (Aa) edge node[auto] {$$} (Fa)
(Fa) edge node[auto] {$$} (rela)
(rela) edge node[auto] {$$} (Aa1)
(Aa1) edge node[auto] {$$} (Fa1)
(Ba) edge node[auto] {$$} (Ga)
(Ga) edge node[auto] {$$} (relb)
(relb) edge node[auto] {$$} (Ba1)
(Ba1) edge node[auto] {$$} (Ga1)
(Aaa) edge node[auto] {$$} (Faa)
(Faa) edge node[auto] {$$} (relaa)
(relaa) edge node[auto] {$$} (Aaa1)
(Aaa1) edge node[auto] {$$} (Faa1)
(Aa) edge node[auto] {$$} (Ba)
(Ba) edge node[auto] {$$} (Aaa)
(Fa) edge node[auto] {$$} (Ga)
(Ga) edge node[auto] {$$} (Faa)
(rela) edge node[auto] {$$} (relb)
(relb) edge node[auto] {$$} (relaa)
(Aa1) edge node[auto] {$$} (Ba1)
(Ba1) edge node[auto] {$$} (Aaa1)
(Fa1) edge node[auto] {$$} (Ga1)
(Ga1) edge node[auto] {$$} (Faa1)
;
\end{tikzpicture}
\caption{Commuting diagrams for Fig. \ref{fig:lemma-eight-2} (d) top path.}
\label{fig:lemma-eight-3}
\end{center}
\end{figure}

\begin{figure}[!ht]
\begin{center}
\begin{tikzpicture}
\matrix[matrix of math nodes,column sep={30pt},row sep={40pt,between origins},nodes={anchor=center}] (s)
{
|[name=Aa]| \Hgroup(A_{\alpha})  & |[name=Fa]| \Hgroup(F_{\alpha})  &|[name=rela]| \Hgroup(F_{\alpha},A_{\alpha})  & |[name=Aa1]| \Hgroup(A_{\alpha})& |[name=Fa1]| \Hgroup(F_{\alpha})\\
|[name=Ba]| \Hgroup(B_{\alpha+\epsilon})  & |[name=Ga]| \Hgroup(G_{\alpha+\epsilon})  &|[name=relb]| \Hgroup(G_{\alpha+\epsilon},B_{\alpha+\epsilon})  & |[name=Ba1]| \Hgroup(B_{\alpha+\epsilon})& |[name=Ga1]| \Hgroup(G_{\alpha+\epsilon})\\
|[name=Aaa]| \Hgroup(B_{\alpha'+\epsilon})  & |[name=Faa]| \Hgroup(G_{\alpha'+\epsilon})  &|[name=relaa]| \Hgroup(G_{\alpha'+\epsilon},B_{\alpha'+\epsilon})  & |[name=Aaa1]| \Hgroup(B_{\alpha'+\epsilon})& |[name=Faa1]| \Hgroup(G_{\alpha'+\epsilon})\\
};
\draw[->] (Aa) edge node[auto] {$$} (Fa)
(Fa) edge node[auto] {$$} (rela)
(rela) edge node[auto] {$$} (Aa1)
(Aa1) edge node[auto] {$$} (Fa1)
(Ba) edge node[auto] {$$} (Ga)
(Ga) edge node[auto] {$$} (relb)
(relb) edge node[auto] {$$} (Ba1)
(Ba1) edge node[auto] {$$} (Ga1)
(Aaa) edge node[auto] {$$} (Faa)
(Faa) edge node[auto] {$$} (relaa)
(relaa) edge node[auto] {$$} (Aaa1)
(Aaa1) edge node[auto] {$$} (Faa1)
(Aa) edge node[auto] {$$} (Ba)
(Ba) edge node[auto] {$$} (Aaa)
(Fa) edge node[auto] {$$} (Ga)
(Ga) edge node[auto] {$$} (Faa)
(rela) edge node[auto] {$$} (relb)
(relb) edge node[auto] {$$} (relaa)
(Aa1) edge node[auto] {$$} (Ba1)
(Ba1) edge node[auto] {$$} (Aaa1)
(Fa1) edge node[auto] {$$} (Ga1)
(Ga1) edge node[auto] {$$} (Faa1)
;
\end{tikzpicture}
\caption{Commuting diagrams for Fig. \ref{fig:lemma-eight-2} (d) bottom path.}
\label{fig:lemma-eight-4}
\end{center}
\end{figure}

This shows that the two commute and hence we conclude that the relative filtrations are interleaved. 
\eop


\subsection{Theorem \ref{theorem:alpha-sampling}}

First we prove a collection of lemmas 
(\ref{lemma:top_filtration}, 
 \ref{lemma:top_filtration_nerve}, 
 \ref{lemma:bottom_filtration_a},
 \ref{lemma:bottom_filtration_b},
 \ref{lemma:nonconvex_proof},
 \ref{lemma:simplicial_interleave}) 
that are relevant in proving Theorem \ref{theorem:alpha-sampling}. 

\begin{lemma}
\label{lemma:top_filtration}
If $\Lspace$ is an $\epsilon$-sample of $\Xspace$ then
$\{\Xspace_\alpha\}$ is $\epsilon$-interleaved with 
$\{\Lspace_\alpha\}$.
\end{lemma}
\proof
Given that  $\Lspace$ is an $\epsilon$-sample of  $\Xspace$, 
by definition, $\Lspace \subseteq \Xspace$, this implies that 
(a) $\Lspace_{\alpha} \subseteq \Xspace_{\alpha}$ and 
(b) $\Lspace_{\alpha+\ep}\subseteq \Xspace_{\alpha+\ep}$.
Subsequently, we would prove by the triangle inequality that,  
(c) $\Xspace_\alpha \subseteq \Lspace_{\alpha+\ep}$.
Combining (a), (b) and (c), we have,
\begin{equation*}
\Lspace_\alpha \subseteq \Xspace_\alpha \subseteq \Lspace_{\alpha+\epsilon}\subseteq \Xspace_{\alpha+\epsilon}.
\end{equation*}
By the special case of $\ep$-interleaving, we have $\Lspace_\alpha \subseteq \Xspace_{\alpha+\epsilon}$ and $\Xspace_\alpha \subseteq \Lspace_{\alpha+\epsilon}$, 
therefore the persistent homology modules of $\{\Xspace_\alpha\}$ and $\{\Lspace_\alpha\}$ is $\ep$-interleaved. 

Now we prove that the inclusion in (c) holds.
For any point $p \in \Xspace_{\alpha}$, 
let $q = \arg\min{_{x \in \Xspace}} d(p, x)$, therefore by definition of $\Xspace_{\alpha}$, 
$d(p, q) \leq \alpha$.
Since $q \in \Xspace$ and $L$ is an $\ep$-sample, 
let $s = \arg\min_{z \in L} d(p,z)$, 
by definition of $L$, $d(q, s) \leq \ep$.
By triangle inequality, $d(p,s) \leq d(p,q) + d(q,s) \leq \alpha + \ep$.
Therefore $p \in \Lspace_{\alpha+\ep}$. 
 \eop

\begin{lemma}
\label{lemma:top_filtration_nerve}
The nerve of $\Lspace_{\alpha}$,
$\Nerve{\Lspace_\alpha}$, is homotopic to $\Lspace_{\alpha}$.
\end{lemma}

\proof
This is an application of the Nerve Theorem. 
Since these are Euclidean balls in Euclidean space, they are all convex as are
all their intersections. They are hence contractible and the
Nerve Theorem applies. 
\eop
\begin{lemma}
\label{lemma:bottom_filtration_a}
 If $\Lspace$ is an  $\ep$-sample of
$\Xspace$, then $\tilde\Lspace$ is a $2\ep$ sample of $\Xspace - \interior{B_r}$. 
\end{lemma}

\proof
Consider a point outside in $\Xspace$
but not in $\interior{B_r}$. If it is covered by a sample lying
outside of $\interior{B_r}$, then it is still with $\epsilon$ of
a sample point. If it is covered by a point within
$\interior{B_r}$, then the closest sample point outside of
$\interior{B_r}$ can be no further than $2\epsilon$. This follows
since all points $\epsilon$ away from the $\interior{B_r}$ cannot
be covered by a sample point which lies within $\interior{B_r}$,
and therefore any point outside $\interior{B_r}$ but covered by a
sample point within $\interior{B_r}$, lies at most $2\epsilon$
from a sample point with lies outside  $\interior{B_r}$ and so is 
in  $\tilde\Lspace$. 

Formally, consider a point $p \in \Xspace - \interior{B_r}$.
Let $s = \arg\min_{z \in L} \d(p,z)$, that is, $p$ is covered by $s$.
If $s$ is outside of $\interior{B_r}$, that is, $s \in \tilde \Lspace$, 
then $d(p,s) \leq \ep$.
If $s \in \interior{B_r}$, 
let $t = \arg\min_{z \in \tilde{L}} d(p, z)$,
we claim that $d(p, t) \leq 2 \ep$. 
Therefore $\tilde{\Lspace}$ is a $2\ep$ sample of $\Xspace - \interior{B_r}$. 
Now we prove the claim that $d(p, t) \leq 2 \ep$.
We could prove by contradiction. 
Suppose $d(p, t) > 2 \ep$ and $p$ is just on the boundary of $\Xspace \cap B_r$.
Then there exists at least a point $z$ that is $\ep$ away from $p$ that is not covered by any sample point in $L$. 
This contradicts with $L$ being an $\ep$-sample.

\eop

\begin{lemma}
\label{lemma:bottom_filtration_b}
$\{\tilde\Lspace_\alpha - \interior{B_r}\}$ is $2\epsilon$-interleaved with
    $\{\Xspace_\alpha-\interior{B_r}\}$. 
\end{lemma}

\proof
The proof follows from the Lemma \ref{lemma:bottom_filtration_a}
and precisely the same argument as in Lemma \ref{lemma:top_filtration}.
\eop


\begin{lemma}
\label{lemma:nonconvex_proof}
For $\alpha<r$, 
the nerve of $\tilde\Lspace_\alpha - \interior{B_r}$ is homotopic
to the union of balls $\tilde\Lspace_\alpha$ with
$\interior{B_r}$ removed.
\end{lemma}
\proof
Since we are removing the ball the intersections are no longer
convex. However the condition $\alpha<r$ ensures that they are
still contractible. This is only an outline of the proof. The
goal is to prove that from any intersection there is a homotopy
to a convex body and hence all the intersections are
contractible. Take an arbitrary intersection. If it does not
intersect $\interior{B_r}$, it is convex. If it does, then take
the tangent plane to $B_r$ at a point on the boundary within the
intersection. Clear the half-plane which does not contain the
$B_r$ intersected with the intersection is convex and hence
contractible.

The rest of the intersection can by retracted to the tangent
plane, which we prove by giving an explicit deformation
retract. The tangent plane will be referred to as $T(s)$ (the
tangent plane at point $s$).

\begin{figure}[htb]
\centering
\subfloat[]{\includegraphics[width=0.25\textwidth,page=1]{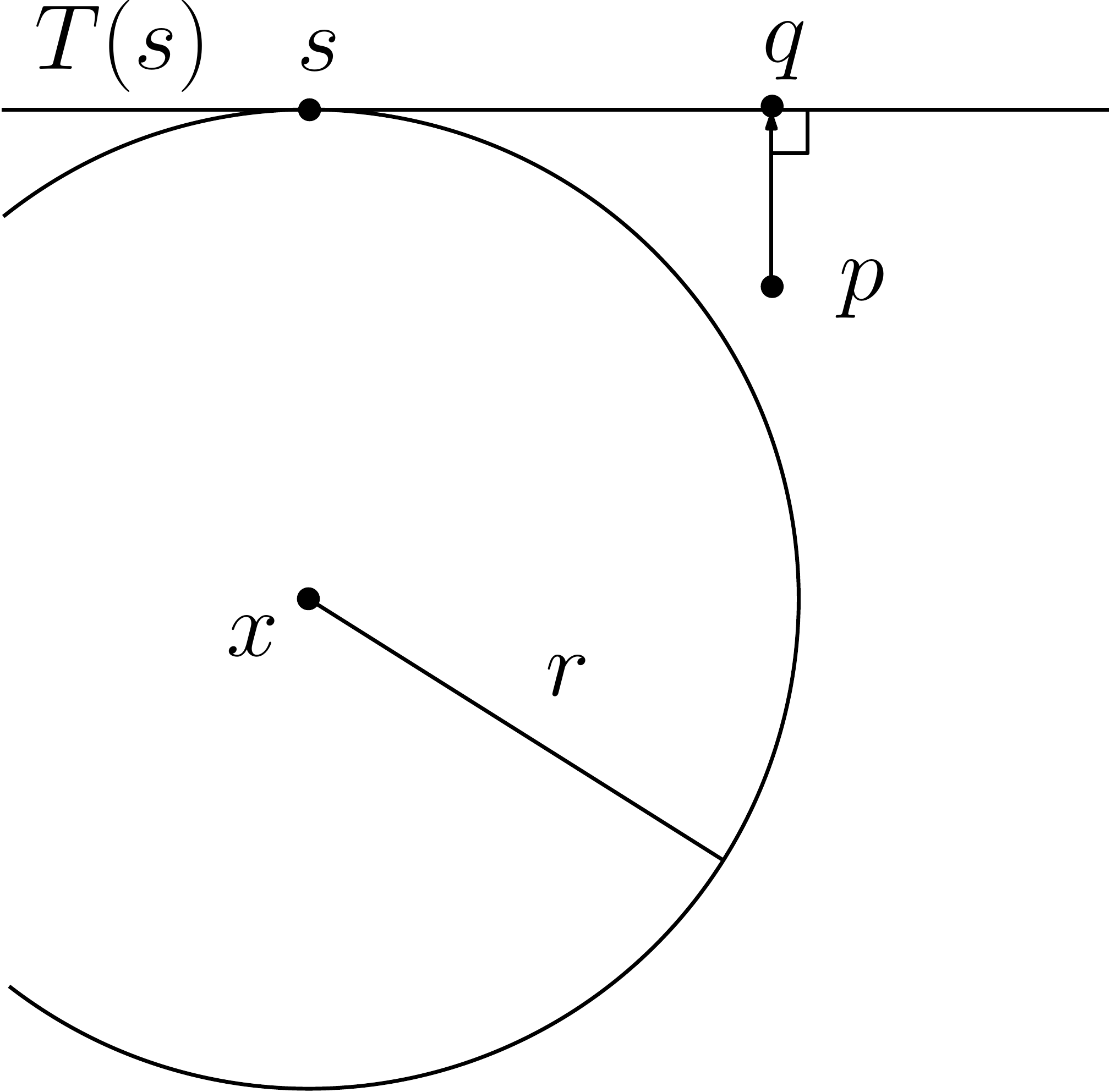}}
\subfloat[]{\includegraphics[width=0.25\textwidth,page=3]{retract}}\\
\subfloat[]{\includegraphics[width=0.25\textwidth,page=4]{retract}}
\subfloat[]{\includegraphics[width=0.25\textwidth,page=5]{retract}}
\caption{\label{fig::retract}(a) The layout of the points $p,q,s$
  along with the deformation retract. (b) The situation when
  $\alpha$ is positive. (c) The situation when $\alpha$ is
  negative (cannot occur). (d) A bound on the distance between
  $p$ and $s$.  Note that although this is in high dimensions,
  these figures are general since we can restrict ourselves to
  the plane defined by $p,q,s$.}
\label{fig:retract}
\end{figure}

First, we define a deformation retract before we remove
$B_r(x)$. We consider a straight-line homotopy to the $T(s)$ by
projection. We project each point $p$ to $T(s)$ \emph{within the
  intersection}. We call this point $q$.  By convexity of the
intersection, this path is a geodesic which lies completely in
the intersection. It is also continuous. With $B_r(x)$ removed
this will remain a valid deformation retract if all geodesics
remain in the space (i.e. pass through $B_r(x)$).  The points
$p$, $q$, $s$ as above shown in Fig. \ref{fig:retract}(a).

  We prove that the geodesic does not leave the space by
 contradiction. Without loss of generality assume the point $p$
 is on the boundary. To leave the space, it must cross the
 boundary of $B_r(x)$, and the shortest path from that point must
 also go through the ball. In particular, we see that to pass
 through $B_r(x)$, the geodesic must form a negative angle with
 the tangent plane, shown by $\alpha$ (Compare Fig. \ref{fig::retract}(b) and
 (c)). Since the line $(p,q)$ is a shortest path to $T(s)$, it
 must be perpendicular to $T(s)$. This implies that the
 angle between $T(p)$ and $T(s)$, denoted by $\beta$, must be
 acute.

This, however implies that the point of contact of the two
hyperplanes is at least $\sqrt{2} r$ far apart as shown in 
Fig. \ref{fig:retract}(d). Since we can choose the point of contact such that no point
on the ball in the intersection is more that $\alpha$ from the
point of contact and $\alpha<r$ (this is obvious if we take
$2\alpha<r$), this implies that such a point cannot be in the
intersection.

Note that the original projection was to $T(s)$ within the
intersection. This means that $(p,q)$ may not be perpendicular to
$T(s)$. However in this case, $(p,q)$ will not go through the
$B_r(x)$. Since $(p,q)$ must form a chord of a ball of radius
$\alpha$, passing through $B_r(x)$ would generate a chord in
$B_r(x)$. This implies that either the center of the ball of
radius $\alpha$ lies within $B_r(x)$ or that $\alpha>r$.

Hence, the projection to $T(s)$ is a deformation retract and the
non-convex part is contractible as well.
\eop

\begin{lemma}
\label{lemma:simplicial_interleave}
$\{\tilde\Lspace_\alpha - \interior{B_r} \}$ is $\left(\frac{\alpha^2}{r}\right)$-interleaved with
    $\{\tilde\Lspace_\alpha\}$.    
\end{lemma}

\proof This proof works at the nerve level. We show that if an
intersection between balls exists in $\tilde\Lspace_\alpha$ it
will exist in $\tilde\Lspace_{\alpha+\alpha^2/r}
-\interior{B_r}$. Clearly any intersection
$\tilde\Lspace_{\alpha} -\interior{B_r}$ is also in
$\tilde\Lspace_\alpha$. If an intersection is in
$\tilde\Lspace_\alpha$ but not in $\tilde\Lspace_{\alpha}
-\interior{B_r}$, this implies it lies in
$\interior{B_r}$. Assuming that there is an intersection
contained within $\interior{B_r}$.  Note that the furthest this
intersection can be from the edge of $\interior{B_r}$ is bounded by
$\alpha^2/r$. The derivation can be found blow. 
Hence the two filtrations are
$(\alpha^2/r)$-interleaved.  

Now we focus on the derivation of $(\alpha^2/r)$-bound. 

\begin{figure}[ht]
\centerline{\includegraphics[width=0.15\textwidth]{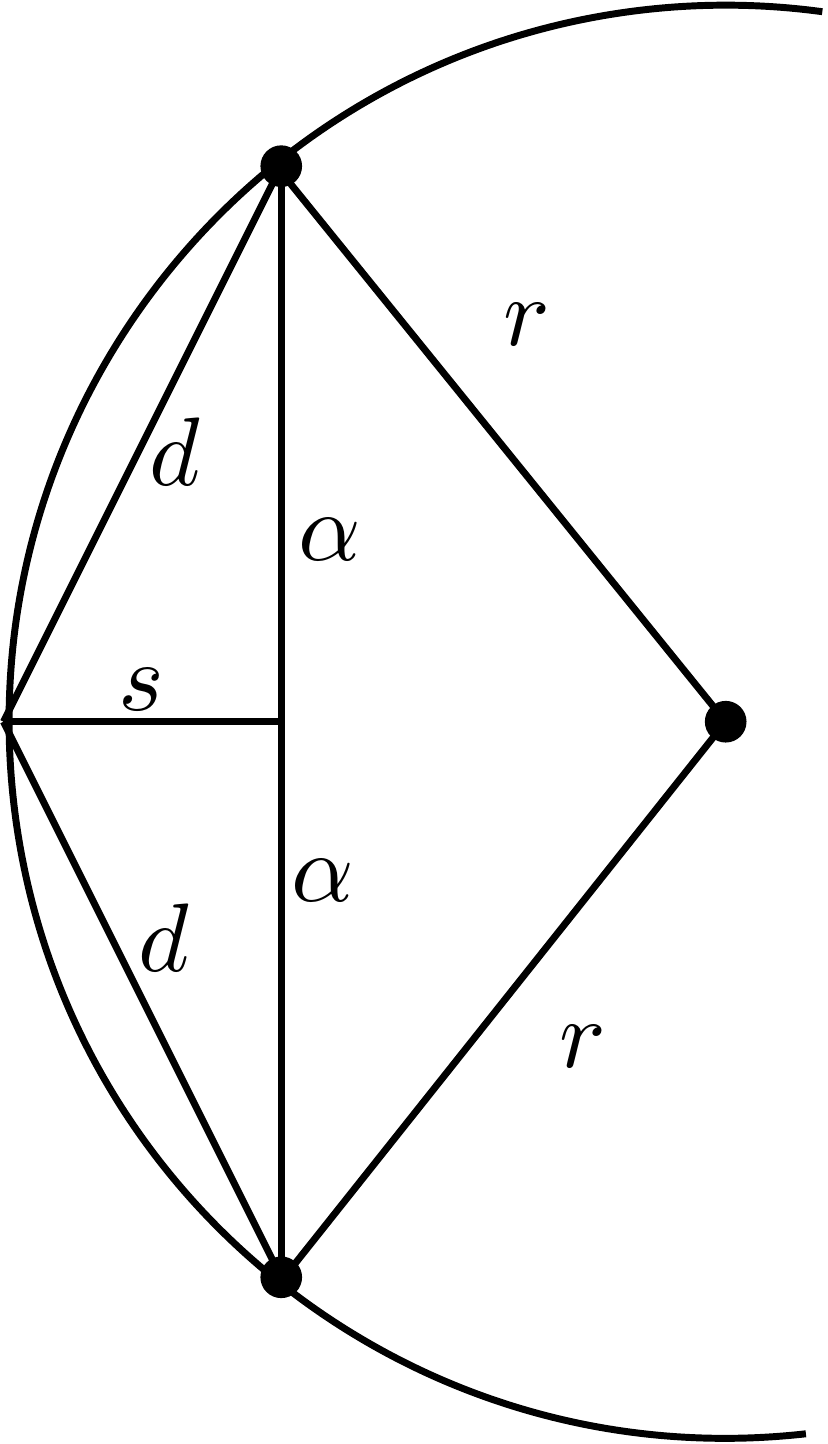}}
\caption{\label{fig:intersection} The geometric situation
  illustrating how deep in the interior of $B_r$ two offsets
  from the exterior of the ball can intersect in terms of the radius of the ball $r$ and  the offset filtration parameter $\alpha$.}
\end{figure}

To prove that the we still get a good approximation we need to
show that if offsets intersect in the ball, they will intersect
soon after outside the ball. The situation is illustrated in
Fig.~\ref{fig:intersection}. Normalizing by $r$, it is a basic
geometric fact that
\begin{equation*}
\frac{s}{r} =  1 - \sqrt{1-\frac{\alpha^2}{r^2}}
\end{equation*}
The distance we must bound, by the Pythagorean theorem is 
\begin{equation*}
\frac{d}{r} = \sqrt{ \left(1 - \sqrt{1-\frac{\alpha^2}{r^2}}\right)^2
  +\frac{\alpha^2}{r^2}}
\end{equation*}
Since $\sqrt{1-\frac{\alpha^2}{r^2}}\geq 1-\frac{\alpha^2}{r^2}$ for $0\leq\frac{\alpha^2}{r^2}\leq 1$ 
\begin{eqnarray*}
\frac{d}{r} &\leq& \sqrt{ \left(1 - 1+\frac{\alpha^2}{r^2}\right)^2
  +\frac{\alpha^2}{r^2}}\\
&=& \sqrt{ \left(\frac{\alpha^2}{r^2}\right)^2
  +\frac{\alpha^2}{r^2}}\\
&=& \sqrt{ \frac{\alpha^4}{r^4}  +\frac{\alpha^2}{r^2}}\leq \frac{\alpha^2}{r^2} +\frac{\alpha}{r}
\end{eqnarray*}

Multiplying by $r$, we see that for $\alpha<r$, we see 
that any simplex (intersection of balls) in $\{\tilde\Lspace_\alpha\}_{\alpha\in [0,\infty)}$ will be in $\{\tilde\Lspace_{\alpha'} - \interior{B_r} \}_{\alpha'\in [0,\infty)}$ for $\alpha +\alpha^2/r \leq \alpha'\leq r $.
We obtained our desired bound. 

\eop


Finally we prove our main theorem for $\alpha$-filtration. 

\noindent\textbf{Theorem \ref{theorem:alpha-sampling}.}
The persistence module with respect to the Vietoris-Rips filtration
of $\{(\Lspace_\alpha,\tilde\Lspace_\alpha)\}$, that is, 
$\{(\Rcal_{\alpha}(\Lspace),\Rcal_{\alpha}(\tilde\Lspace))\}$ is 
$\left(2\epsilon +\alpha + \frac{\alpha^2}{r}\right)$-interleaved with 
the $\alpha$-filtration, $\{(\Xspace_{\alpha}, \Xspace_{\alpha} - \interior B_r)\}$,  for $\alpha<r$.
\proof 
Lemma \ref{lemma:top_filtration} tells us $\{\Xspace_{\alpha}\}$ and $\{L_{\alpha}\}$ are $\ep$-interleaved. 
Lemma \ref{lemma:top_filtration_nerve} shows $\Nerve{L_{\alpha}} \simeq L_{\alpha}$.
Lemma \ref{lemma:bottom_filtration_b} states $\{\Xspace_{\alpha} - \interior B_r\}$ and $\{\tilde{L}_{\alpha} - \interior B_r\}$ are $2\ep$-interleaved. 
Lemma \ref{lemma:nonconvex_proof} shows $\Nerve{\tilde{L}_{\alpha} - \interior B_r} \simeq \tilde{L}_{\alpha} - \interior B_r$.
Lemma \ref{lemma:simplicial_interleave} indicates $\{\tilde{L}_{\alpha} - \interior B_r\}$ and $\{\tilde\Lspace_\alpha\}$ are $\frac{\alpha^2}{r}$-interleaved. 

Lemma \ref{lemma:bottom_filtration_b} and \ref{lemma:simplicial_interleave} implies that $\{\Xspace_{\alpha} - \interior B_r\}$ and $\{\tilde\Lspace_\alpha\}$ are $(2\ep+\frac{\alpha^2}{r})$-interleaved. 
Combined with Lemma \ref{lemma:top_filtration}, we have the relative modules, 
$\{(\Xspace_{\alpha}, \Xspace_{\alpha} - \interior B_r)\}$ 
and $\{(L_{\alpha}, \tilde\Lspace_\alpha)\}$ are $(2\ep+\frac{\alpha^2}{r})$-interleaved. 
This means, the persistence diagram of the \v{C}ech filtration of $\{(\Lspace_\alpha,\tilde\Lspace_\alpha)\}$ 
is $(2\ep+\frac{\alpha^2}{r})$ approximation of the persistence diagram of $\alpha$-filtration. 

Now we consider \v{C}ech filtrations for both $\{L_{\alpha}\}$ and $\{\tilde\Lspace_\alpha\}$, that is, 
$\{\Ccal_{\alpha}(\Lspace)\}$ and $\{\Ccal_{\alpha}(\tilde{\Lspace})\}$. 
Since both are $\alpha$-interleaved with their Vietoris-Rips counterparts, 
that is, $\{\Ccal_{\alpha}(\Lspace)\}$ is $\alpha$-interleaved with $\{\Rcal_{\alpha}(\Lspace)\}$, 
and $\{\Ccal_{\alpha}(\tilde{\Lspace})\}$ is $\alpha$-interleaved with $\{\Rcal_{\alpha}(\tilde\Lspace)\}$. 
We lose a factor $\alpha$ in the approximation by switching to Vietoris-Rips filtration of $\{(\Lspace_\alpha,\tilde\Lspace_\alpha)\}$. 
\eop


\section{Discussion on the  $r$-filtration}
\label{sec-app:proofs-r}

\begin{figure}
\vspace{-5mm}
\centering\includegraphics[width=0.3\textwidth]{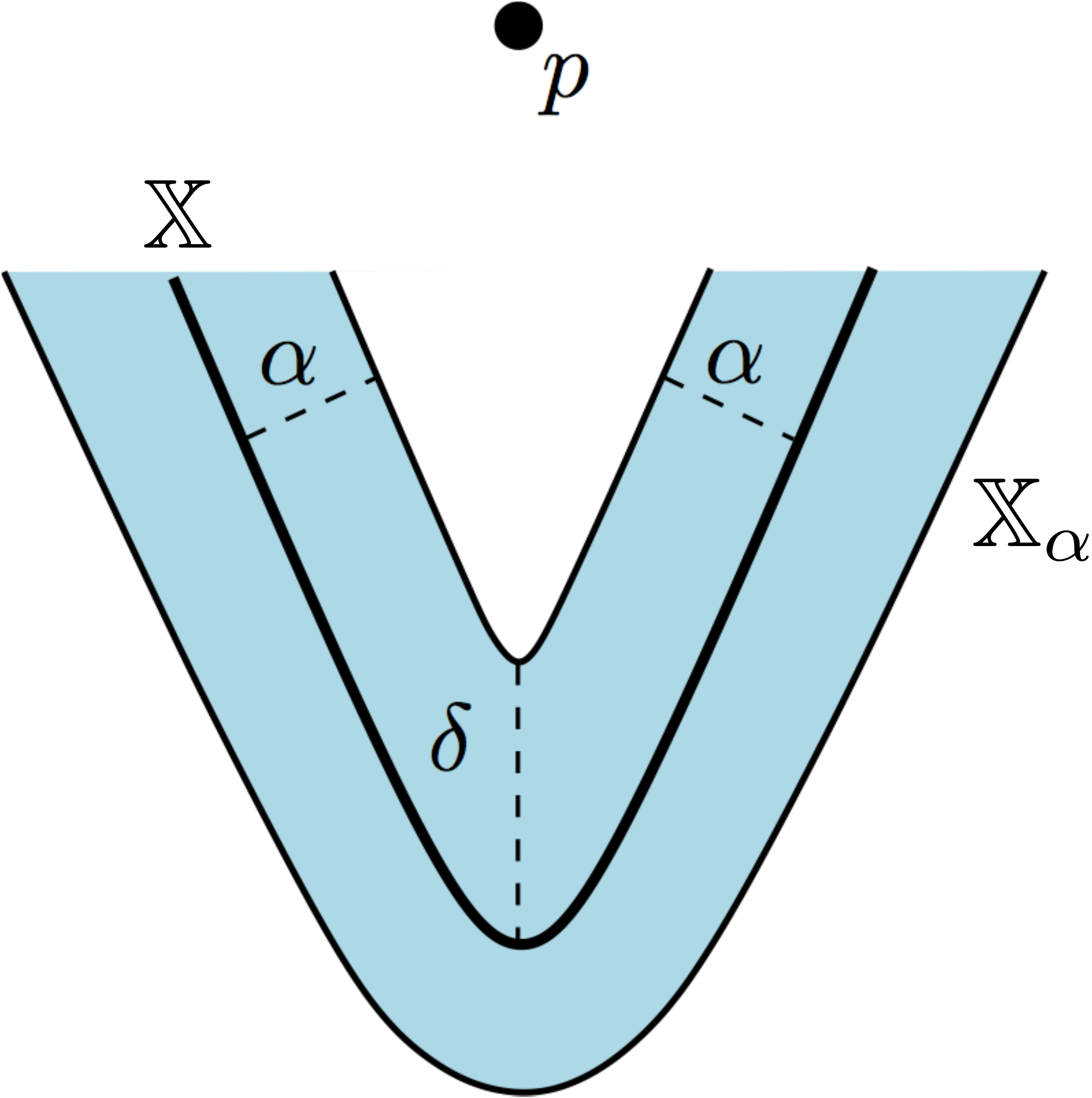}
\caption{\label{fig:badcase} The space $\Xspace$, its $\alpha$-offset $\Xspace_{\alpha}$
  and a point $p$. Consider the distance function $d_p$ to $p$. This is
  an example where the offset $\Xspace_{\alpha}$ and space $\Xspace$ are homotopy equivalent but the
  persistence diagrams of the functions $\restr{d_p}{\Xspace}$ and $\restr{d_p}{\Xspace_{\alpha}}$ are potentially far apart. }
\vspace{-5mm}
\end{figure}

Here we give a short discussion on the assumptions
made in the Section~\ref{sec:r_filt} and relate it to existing work. Our
primary assumption is an $\ep$-homotopy equivalence between a pair
of spaces. This is a strong assumption since it requires that
points are only moved a bounded amount in the homotopy,
essentially excluding situations illustrated in Fig.~\ref{fig:badcase}.
Here we define the Euclidean distance function to the point $p$ as 
$d_p(x) := d(p,x) = ||x-p||$. 
Now consider $d_p$ restricted to $\Xspace$ and $\Xspace_{\alpha}$.  
Although $\Xspace$ could be approximated via a retract from $\Xspace_{\alpha}$, 
it is insufficient to guarantee that we could well-approximate the persistence module 
of $\restr{d_p}{\Xspace}$ through that of $\restr{d_p}{\Xspace_{\alpha}}$.  
In other words, in Fig.~\ref{fig:badcase}, the persistence diagram of these persistence modules 
could differ by at least $\delta$. 


The problem of approximating a sublevel set filtration of a function on a space
has been studied before. The setting is closely related to the
results of ~\cite{ChazalGuibasOudot2009}.  In ~\cite{ChazalGuibasOudot2009},
there is an approximation guarantee between a sublevel set
filtration of a $c$-Lipschitz function on a space and an image
persistence filtration on two nested Vietoris-Rips complexes with an
appropriately chosen parameter. There are numerous requirements
to apply such results, which we outline here.

The first requirement is that we have access to geodesic
distances or some provable approximation of it. While the
geodesic distance can be inferred from the Euclidean distance in
certain cases, this can be a difficult problem depending on how
our space is embedded. The second requirement is that the space
has positive convexity radius. While this is generally a safe
assumption for manifolds; for spaces where local homology yields
interesting information, such as stratified spaces,  this
measure can often be zero (i.e. a cone has zero convexity
radius). If, however such requirement is satisfied, we can apply the
results in \cite{ChazalGuibasOudot2009} directly.  
The resulting algorithm is to build the
underlying simplicial complex using geodesic distances, which
given a sufficiently dense sampling relative to the convexity
radius, gives an approximation for \emph{any} $c$-Lipschitz
function. Since distance functions are 1-Lipschitz, the
approximation results follow.

This highlights a key obstacle in stating sampling results for
function filtrations as well as an open problem we discuss below:  
in terms of which measures should we state
sampling results? Is there a global geometric measure which is meaningful for stratified spaces?
Are there weaker conditions than $\ep$-homotopy for approximating sublevel set behavior?
As pointed out above, geometric measures, 
such as \emph{reach} or \emph{convexity radius} can be zero even for nice
spaces. 
It would be
preferable to use quantifiers such as \emph{homological feature size}~\cite{CohEdeHar2007}.
Research in these directions is left for future work.

\end{document}